\documentclass[lettersize,journal]{IEEEtran}
\usepackage{amsmath,amsfonts,amsthm}
\usepackage{algorithmic}
\usepackage{algorithm}
\usepackage{array}
\usepackage[caption=false,font=normalsize,labelfont=sf,textfont=sf]{subfig}
\usepackage{textcomp}
\usepackage{stfloats}
\usepackage{url}
\usepackage{verbatim}
\usepackage{cite}
\usepackage[autostyle, english = american]{csquotes}
\usepackage[english]{babel}
\usepackage{url}
\usepackage{bbm}
\usepackage{tfrupee}
\usepackage{enumitem}
\usepackage[hidelinks, colorlinks=true, allcolors=blue]{hyperref}
\usepackage{amssymb}
\usepackage{pifont}
\usepackage{color}
\usepackage{xcolor}
\usepackage{nomencl}
\usepackage{autobreak}
\usepackage{mathtools}
\usepackage{bm}

\newtheorem{prop}{Proposition}
\newtheorem{definition}{Definition}
\newtheorem{remark}{Remark}
\makenomenclature
\renewcommand{\nomgroup}[1]{%
  \item[
    \ifthenelse{\equal{#1}{A}}{\textit{Acronyms and Parameters}}{%
    \ifthenelse{\equal{#1}{V}}{\textit{Variables}}{%
    \ifthenelse{\equal{#1}{S}}{\textit{Sets}}{}}}%
  ]%
}

\definecolor{darkgreen}{rgb}{0.0, 0.5, 0.0}
\definecolor{darkred}{rgb}{0.5, 0.0, 0.0}
\begin{document}
% \title{Design of IDSO-Managed Bid-Based Transactive Distribution Systems: DER Participation in Wholesale Markets with Preserved T–D Interactions}
\title{IDSO-Managed Bid-Based Transactive Distribution Systems\\Design for DER Participation in Wholesale Markets\\While Preserving T–D Interactions}
% \title{Design of IDSO-Managed Bid-Based Transactive Distribution System: Aggregating DERs for Wholesale Markets While Preserving T–D Interactions}
\author{Swastik Sharma,~\IEEEmembership{Student Member,~IEEE,} Swathi Battula,~\IEEEmembership{Member,~IEEE}, and Sri Niwas Singh,~\IEEEmembership{Fellow,~IEEE}\vspace{-2em}
        % <-this % stops a space
}
% The paper headers
\markboth{}%
{Shell \MakeLowercase{\textit{et al.}}: A Sample Article Using IEEEtran.cls for IEEE Journals}
% \IEEEpubid{0000--0000/00\$00.00~\copyright~2025 IEEE}
% Remember, if you use this you must call \IEEEpubidadjcol in the second
% column for its text to clear the IEEEpubid mark.
\twocolumn[%
\begin{@twocolumnfalse}
\thispagestyle{empty} % removes page number
This work has been submitted to the IEEE for possible publication. Copyright may be transferred without notice, after which this version may no longer be accessible.
\vspace{1cm} % adjust space before switching to double-column mode
\end{@twocolumnfalse}
]
\maketitle
\begin{abstract} 
Participation of Distributed Energy Resources (DERs) in bid-based Transactive Energy Systems (TES) at the distribution systems facilitates strongly coupled, bidirectional interactions between Transmission-Distribution (T-D) systems. Capturing these interactions is critical for ensuring seamless integration within an Integrated Transmission and Distribution (ITD) framework. This study proposes a methodology to preserve such tight T-D linkages by developing an Independent Distribution System Operator (IDSO) managed bid-based TES design for unbalanced distribution systems. The proposed design operates within the ITD paradigm and permits DER participation in the Wholesale Power Market (WPM) through IDSO while preserving tight T-D linkages. To this end, this research offers the following key contributions: a novel bid/offer prequalification-cum-aggregation method to ensure a grid-safe and value-based aggregation of DERs' bids and offers for WPM participation through IDSO; and a retail pricing mechanism that reflects the true value of procuring or offering additional units of power within the distribution system. Case studies are conducted on a modified IEEE 123-bus radial feeder populated with a high DER concentration to validate the proposed frameworks' effectiveness in coordinating the DERs efficiently and reliably.
\end{abstract}
\vspace{-0.9em}
\section{Introduction}
\subsection{Motivation}
\IEEEPARstart{F}{ederal} Energy Regulatory Commission (FERC) Order 2222 \cite{ferc2020} provides an opportunity for Distributed Energy Resources (DERs) to participate in the Wholesale Power Market (WPM) through aggregation. DERs' participation allows the Independent Transmission System Operator (ITSO) to leverage their flexibility and enhance grid reliability under uncertainty. However, it is essential to value this flexibility--not only from the perspective of system operators but also from the perspective of DERs, as aligning these differing vantage points is crucial for reliable coordination and market integration of these resources.
\par To facilitate this dynamic value exchange and enhance coordination among diverse actors, Transactive Energy System (TES)-based distribution system designs \cite{GridWiseTES} are gaining momentum. These designs empower customers with greater autonomy by exposing them to market-driven prices while simultaneously exploiting their price reactivity to provide price-based control through bidirectional communication \cite{Battula_TES, SharmaTES}. Recently, researchers have started exploring bid-based TES designs employing an Independent Distribution System Operator (IDSO) \cite{Battula_TES, PNNL_5}. Such a design lets DERs express their goals and constraints through price-volume-based bids and offers, alleviating privacy concerns. Meanwhile, the IDSO, acting as an aggregator, manages the DERs and maintains the network. Although promising, the mechanisms must ensure distribution network reliability while performing value-based aggregation for WPM participation to ensure seamless interactions within Integrated Transmission and Distribution (ITD) Systems.
\vspace{-0.9em}
\subsection{Literature Review}
The seminal work on TES designs is credited to \textit{Pacific Northwest National Laboratories (PNNL)}, whose demonstrations, as reported in \cite{PNNL_3, PNNL_4}, have proven the merits of such a design. For an in-depth analysis and review of research on TES designs, see \cite{Renani}. TES designs have since diversified into several categories, including Local Energy Market (LEM)-based approaches \cite{Ranjbar_TES, Saber3_TES, Bedoya_BM, Zhou_P2P, Mukh_TES, Affolabi_TES}, consensus-based models \cite{Xiao_TES, Hu_TES, Rui_tesfatsi}, bid-based frameworks \cite{Battula_TES, Nazir_TES}, and hybrid methods \cite{Ullah_TES_TSO, Saber_TES, Hoque_TES, Hoque2_TES} that combine elements of each.

LEM-based approaches focus on integrating DERs into small-scale retail or bilateral markets on the distribution side, allowing them to provide services to the distribution network. For example, as in \cite{Ranjbar_TES, Saber3_TES, Bedoya_BM}, the DERs submit bids and offers in an LEM-based TES framework for procuring power and offering services. However, none of these studies explores the aggregation of DERs for participation in WPM while ensuring the reliability of the distribution network. 
\par Consensus-based models establish retail price sequences for scheduling DERs by facilitating information exchange between stakeholders through negotiation. Such as \cite{Hu_TES} introduces a DSO-based TES design with a price coordinator mediating between the DSO and aggregators, while \cite{Rui_tesfatsi} presents an IDSO-managed consensus framework for unbalanced networks. 
Although both \cite{Hu_TES} and \cite{Rui_tesfatsi} consider distribution network constraints, they arrive at retail price signals through negotiations rather than through WPM participation of DERs. Hence, these models may not accurately reflect the true value of procuring or offering additional units of power and may also face convergence issues in real-world applications.
\par Bid-based frameworks use bids/offers that are price-volume pairs indicating their reservation values.
In \cite{Battula_TES}, the authors derive state-conditioned bid/offer prices for a smart HVAC considering welfare attributes and local constraints. Nazir and Hiskens in \cite{Nazir_TES} develop a bin-based DER aggregate model that uses user-defined bid functions to address issues like load synchronisation. While bid-based frameworks inherently enable value-based aggregation of DERs and thus have the potential to support WPM participation, the authors in \cite{Battula_TES, Nazir_TES} have not fully explored this application, which requires explicitly modelling the distribution network constraints into the DSO's optimization formulation.
\par Hybrid designs merge features from previous frameworks for a more comprehensive DER coordination. For instance, \cite{Ullah_TES_TSO} proposes a decentralised TE market that connects wholesale and distribution-level markets through interactions among TSOs, DSOs, and DERs. It uses a cost-minimisation approach with consensus-based negotiations, employing iterative price signals. In \cite{Saber_TES, Hoque_TES, Hoque2_TES}, a bid-based framework is combined with an LEM design where EVs receive DLMPs from the DSO, generate bids/offers, and feed them back to create a charging schedule that meets system objectives. However, these studies do not investigate value-based DER aggregation for WPM participation that ensures network reliability.

Furthermore, IDSO-managed TES designs at the distribution systems enable tight bidirectional exchanges of power, prices and data at the T-D interface \cite{ITDPlatform}. 
These exchanges that occur between ITSO and IDSO---involving the aggregated bid/offer functions and WPM outcomes---affect the exchanges that occur between the IDSO and DERs---involving retail signals and communication of bids and offers, thereby influencing the behaviour of the ITD system under high DER penetration. Prior studies (see \cite{Ranjbar_TES, Affolabi_TES, Saber_TES, Hu_TES, Rui_tesfatsi, Battula_TES, Hoque_TES, Hoque2_TES}) consider the power procurement of DSO assuming either the WPM Locational Marginal Price (LMP) or the net demand is known, treating them as exogenous variables. This eliminates the feedback that the two-way linkages offer. The DER net-load consumption and the WPM LMP exhibit interdependence; neither can be treated as constant while simulating tight two-way linkages and their mutual impacts. To tackle this, the studies mostly use either a large-scale integrated optimisation problem \cite{Largescale} or iterative methods \cite{Ullah_TES_TSO, iterative2} to arrive at converged LMP and aggregated distribution net demand. However, the former methods rely on complete knowledge of both systems, making assumptions that fail to capture the dynamics of real ITD systems. The latter methods are computationally intensive with the growing size of systems and time horizons.
\vspace{-0.9em}
\subsection{Research Gaps} 
The following major research gaps are identified:

\begin{itemize}
    \item Lack of TES designs that perform both value-based and grid-safe DER aggregation to enable their WPM participation through aggregated bid/offer value functions and support open-ended dynamic WPM operations. 
    \item Absence of methods that determine retail prices which reflect the true value of procuring or offering additional units of power within the distribution systems, while also incorporating the network-associated costs.
    \item Need for TES designs that permit feedback based on tight T-D linkages to ensure aggregation of DERs at the distribution system supports overall efficiency and reliability of ITD systems. 
\end{itemize}
\vspace{-0.9em}
\subsection{Contributions and Paper Organisation}
This research proposes an IDSO-managed bid-based TES design of unbalanced distribution networks that performs a value-based and grid-safe aggregation of DERs to enable their participation in WPM. The IDSO's resultant aggregated bid/offer functions support tight two-way linkages of the T-D interface.
The key attributes of this study are---
\begin{itemize}
    \item A novel bid/offer prequalification-cum-aggregation method based on Three-Phase Unbalanced Distribution OPF (T-DOPF) is proposed. The proposed aggregation problem performs grid-safe and value-based aggregation of the DERs to participate in WPM while ensuring recovery of network costs.
    \item A shadow price-based retail pricing design is proposed, which reflects the true value of procuring/offering additional units of power within the distribution system. As a result, these signals enable them to update their bids and offers for subsequent time periods.
    \item The proposed design enables tight two-way T-D linkages between IDSO and ITSO by treating the distribution system net demand and WPM LMP as unknown variables. 
\end{itemize}
\par The paper is organized as follows.
Section \ref{sec:overview} provides the general features of the TES design employed in this paper. Section \ref{sec:DN} discusses the power modelling in an unbalanced distribution network. 
The modelling of IDSO and the proposed methodology of this work are discussed in Section \ref{sec:model} and Section \ref{sec:methodology}, respectively. Section \ref{sec:Results} reports test cases that validate the efficacy of the proposed design, and Section \ref{sec:conclusion} concludes the findings of this work. A quick-reference nomenclature is provided in Appendix~A.
\section{Bid-Based TES Design for Unbalanced Distribution Networks}
\label{sec:overview}
\begin{figure}[!htbp]
    \centering
    \includegraphics[width=\linewidth]{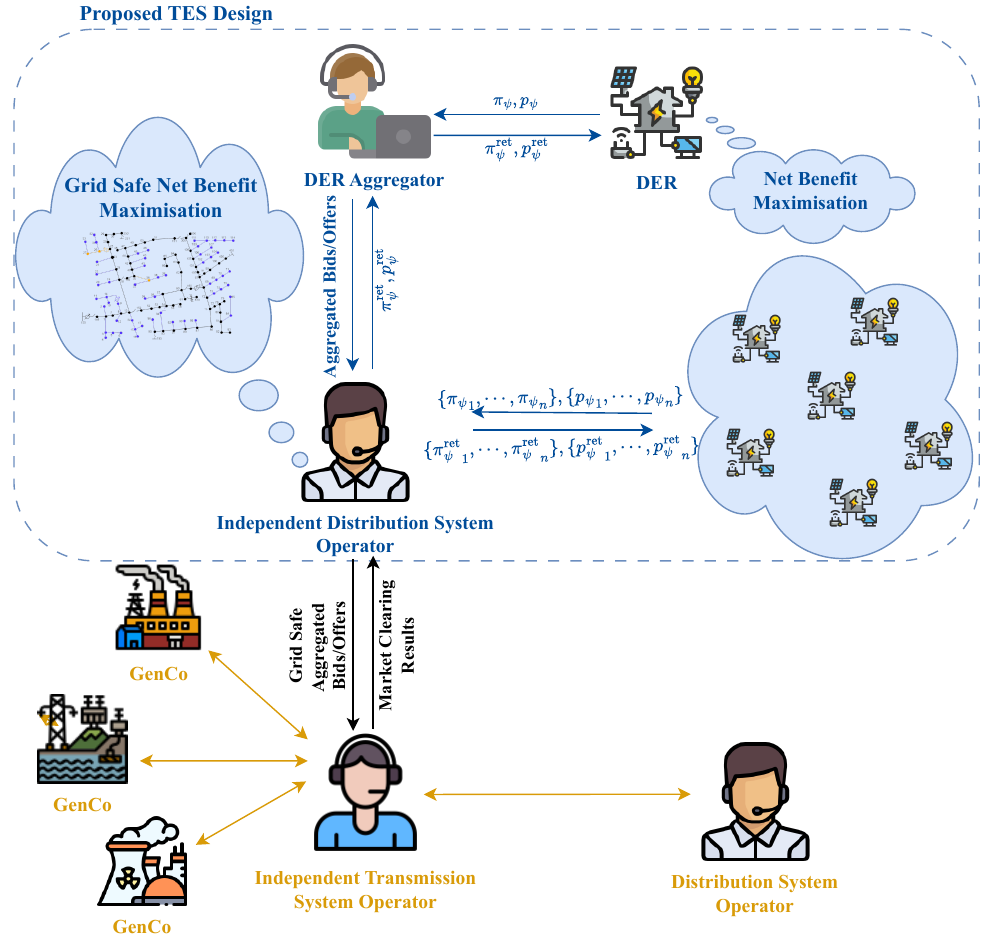}
    \caption{An IDSO-managed bid-based TES design for distribution systems operating within an ITD system.}
    \label{fig:heirarchical}
\end{figure}

The general features of the proposed bid-based TES design are briefly described in this section. The proposed bid-based TES design for an unbalanced distribution network, as shown in Fig. \ref{fig:heirarchical}, is considered to be operating within an ITD system. 
As shown in the figure, the IDSO serves as the link between transmission and distribution systems, facilitating coordinated interactions both with DERs and the ITSO. The ITSO operates the WPM in the transmission system, where bulk Generating Companies (GenCos) and loads participate, and determines WPM LMPs. This study considers IDSO’s participation only in the Real-Time Market (RTM) operated by the ITSO. The information exchange among these entities at the T-D interface includes bid/offer functions (IDSO to ITSO) and RTM LMPs (ITSO to IDSO). 
\par At the distribution side, the DERs arrive at price-quantity pairs of bids/offers and communicate this to the IDSO. The IDSO participates in the WPM by performing a grid-safe and value-based aggregation of these bids/offers by solving the proposed aggregation problem described in Section~\ref{sec:model}. For the DERs or DERAs cleared in the WPM, the IDSO charges/pays only for their active power consumption/production. The IDSO then determines and communicates value-driven retail price signals to all participating DERs. For those DERs with bids/offers that are not cleared, these signals indicate the price that would have ensured their selection during the aggregation and subsequently in WPM.
\par
\begin{figure}[!htbp]
    \centering
    \includegraphics[width=\linewidth]{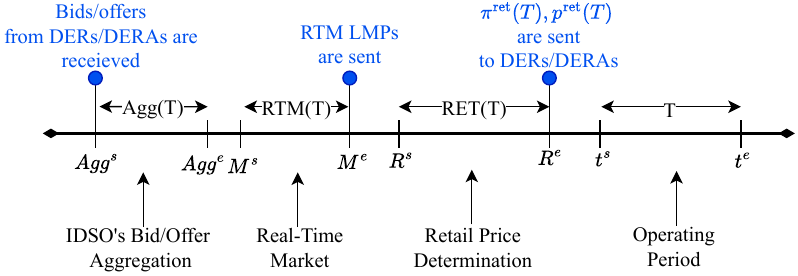}
    \caption{Timing configuration of bid-based TES design to that of the WPM.}
    \label{fig:timing}
    \vspace{-2em}
\end{figure}

 The timing configuration of the proposed design is shown in Fig. \ref{fig:timing}. 
For an operating period T, before the WPM operations commence, the IDSO engages in the grid-safe and value-based aggregation during the period Agg(T)\nomenclature[A]{Agg(T)}{IDSO's aggregation period for operating period T;}. DERs send in their bids/offers before the beginning of Agg(T), and the IDSO performs \textit{bid/offer prequalification-cum-aggregation} by the end of Agg(T). Using the qualified bids/offers, it participates in the RTM\nomenclature[A]{RTM(T)}{Real Time Market operation period for an operating period (T);}. The ITSO clears the market and announces the market outcomes. Using the market outcomes, the IDSO determines retail signals, $\pi^\mathrm{ret}(T),p^\mathrm{ret}(T)$\nomenclature[A]{\(\pi^\mathrm{ret}(T),p^\mathrm{ret}(T)\)}{Retail price (\textcent/kWh) and volume (kW) signals to be sent to DERs for operating period T;} for the operating period T during RET(T)\nomenclature[A]{RET(T)}{Retail price determination period for operating period T;} and sends it to the participating DERs and DERAs before the start of the operating period T. This study assumes the IDSO is the only aggregator considered for the distribution system, in direct contact with the DERs.
\section{Distribution Network and DER Model}
\label{sec:DN}
\subsection{Distribution Network Model}
This study considers an unbalanced radial distribution network with $N+1$\nomenclature[A]{\(N\)}{Number of non-head buses in distribution network;} buses and unbalanced\nomenclature[S]{\(\mathcal{N}\)}{Set containing all the non-head buses of distribution system;}\nomenclature[S]{\(\mathcal{L}\)}{Set containing the N distinct line segments of distribution system;}phases $\phi \in \bm{\Phi}: \{a,b,c\}$\nomenclature[A]{\(\phi\)}{Phase connection of a line, bus or DER;}\nomenclature[S]{\(\bm{\Phi}\)}{Set of valid phases;}. The index of the head bus in the feeder where the IDSO's substation is located is denoted by \nomenclature[A]{\(0\)}{Head bus of the radial network at which IDSO's substation is located, linking Transmission and Distribution systems;}$0$, and all other non-head buses are the elements of the set $\mathcal{N}: \{1,2, \cdots, N\}$. The distribution network has $N$ distinct line segments denoted by set $\mathcal{L} \subseteq \{\mathcal{N} \cup \{0\}\} \times \{\mathcal{N} \cup \{0\}\}$, with $(i,j) \in \mathcal{L}$\nomenclature[A]{\((i,j)\)}{Line segment connecting buses \(i \ \text{and } j\);} denoting a line segment between buses $i$ and $j$. 
This paper adopts the three-phase \textit{LinDistFlow} model proposed by Rui et al. in \cite{Rui_tesfatsi}, elaborated in Appendix~\ref{Append:DistFlowModel}. The power flow equations, reproduced verbatim from Appendix~\ref{Append:DistFlowModel}, are presented below:
\begin{subequations}
% \label{eqn:MatrixLinDistFlowFinal}
    \begin{align}
    % \label{eqn:eqn:MatrixLinDistFlowFinalp0q0}
    \bm{c}^T_0\bm{P} = \bm{p}_0; &\quad \bm{c}^T_0\bm{Q} = \bm{q}_0;\\
    \label{eq:MatrixLinDistFlowFinalpq}
    \bm{C}^T\bm{P} = \bm{p}_{1:N}; &\quad \bm{C}^T\bm{Q} = \bm{q}_{1:N};\\
    % \label{eqn:MatrixLinDistFlowVFinal2}
        \bm{v}_{1:N}
        = \bm{I}^3_N\bm{v}_0 + &2 \bm{C}^{-1} \left[ \bm{D_r} \bm{P} + \bm{D_x} \bm{Q} \right]. 
    \end{align}
\end{subequations}Here $\bm{c}_0 \in \{-\bm{I}_3,0,\bm{I}_3\}^{3N \times 3}$ is the column of the branch–bus incidence matrix, $\bm{\tilde{C}} = [\bm{c}_0, \bm{C}]$, that represents the head bus $0$, and $\bm{C} \in \{-\bm{I}_3,0,\bm{I}_3\}^{3N \times 3N}$ represents the remaining buses. The column vectors for three-phase real and reactive power injections are denoted as $\bm{p}_0,\bm{q}_0 \in \mathbb{R}^{3 \times 1}$ for the head bus, and $\bm{p}_{1:N},\bm{q}_{1:N} \in \mathbb{R}^{3N \times 1}$ for the non-head buses. Vectors $\bm{P}$ and $\bm{Q} \in \mathbb{R}^{3N \times 1}$ represent the three-phase real and reactive line flows across all N lines, respectively. Voltages are stored as squared magnitudes, defined as $\bm{v}_0 \in \mathbb{R}^{3 \times 1}$ for head bus and $\bm{v}_{1:N} \in \mathbb{R}^{3N \times 1}$ for non-head buses, respectively. Finally, $ \bm{D_r}, \bm{D}_x \in \mathbb{R}^{3N \times 3N}$ are diagonal matrices whose $3 \times 3$ main diagonal blocks hold the phase-coupled resistances ($\bar{\bm{R}}_{ij}$) and reactances ($\bar{\bm{X}}_{ij}$) of each line segment ($i,j$) and all off-diagonal blocks are zero matrices.\nomenclature[A]{\(\bm{C}\)}{Branch-bus incidence matrix excluding the head bus;}\nomenclature[A]{\(\bm{c}_0\)}{Branch-bus incidence matrix of the head bus;}\nomenclature[V]{\(\bm{P, Q}\)}{3-phase real and reactive power flow (p.u.) column vector across all lines in the distribution network;}\nomenclature[V]{\(\bm{p}_0, \bm{q}_0\)}{3-phase real and reactive power injection (p.u.) column vector of the head bus;}\nomenclature[V]{\(\bm{p}_{1:N}, \bm{q}_{1:N}\)}{3-phase real and reactive power injection (p.u.) column vector at all non-head buses;}\nomenclature[V]{\(\bm{v}_0\)}{3-phase squared voltage magnitude of head bus in p.u.;}\nomenclature[V]{\(\bm{v}_{1:N}\)}{3-phase squared voltage magnitude (p.u.) of all non-head buses;}\nomenclature[A]{\(\bm{D_r}, \bm{D_x}\)}{Diagonal matrices whose $3 \times 3$ main diagonal blocks hold the phase-coupled resistances ($\bar{\bm{R}}_{ij}$) and reactances ($\bar{\bm{X}}_{ij}$) of each line segment ($i,j$) and all off-diagonal blocks are zero matrices.}
\vspace{-0.9em}
\subsection{Distributed Energy Resource Model}
Let $\bm{\Psi}$\nomenclature[S]{\(\bm{\Psi}\)}{Set denoting all the DERs;} represent the set of DERs participating in the bid-based TES framework, with $\bm{\Psi}_i$\nomenclature[S]{\(\bm{\Psi}_i\)}{Set denoting all the DERs connected at bus $i$;} denoting the subset of DERs connected to bus $i$. The sets of DERs with bids and offers are denoted by \(\bm{\Psi}^-\) and \(\bm{\Psi}^+\), respectively\nomenclature[S]{\(\bm{\Psi}^+\)}{Set of DERs with offers;}\nomenclature[S]{\(\bm{\Psi}^-\)}{Set of DERs with bids;}. Each DER $\psi$\nomenclature[S]{\(\psi\)}{An individual DER, \(\psi \in \bm{\Psi}\);} is characterized by the attribute tuple $(\pi(\psi), p(\psi), \bm{\Phi}(\psi), \theta(\psi), i)$\nomenclature[A]{\(\pi(\psi), p(\psi)\)}{Bid/offer price (\textcent/kWh) and volume (kW) of a DER \(\psi \in \bm{\Psi}\);}\nomenclature[A]{\(\bm{\Phi}(\psi) \subset \bm{\Phi}\)}{Set containing phase connection of a DER $ \psi$;}\nomenclature[A]{\(\theta(\psi)\)}{Power factor in \((0,1]\) (unit-less) for a DER \(\psi\)'s operation;}. At the start of the IDSO's aggregation process, DERs send their price-volume bids/offers specified by $\pi(\psi)$ (¢/kWh) and $p(\psi)$ (kW), respectively. This work adopts the convention, $\pi(\psi) \geq 0$, which applies to both bids and offers, where $p(\psi) < 0$ represents bids and $p(\psi) > 0$ represents offers. DERs are connected to the distribution network via single-phase, two-phase, or three-phase topologies, with $\bm{\Phi}(\psi) \subset \bm{\Phi}$, a set indicating their phase connections. The power factor of operation is $\theta(\psi)$ (unit-less), and $i$ specifies the connected bus.
\par This research takes the liberty of making a simplifying assumption that the per-phase injection of a DER, $\psi$, in p.u. is given by 
\begin{equation}
    \label{eq:PerPhase}
    \begin{aligned}
        p^\phi(\psi) = \begin{cases}
           \displaystyle \frac{p(\psi)}{S_\mathrm{base}N_{\phi_\psi}}; &\text{ if } \phi \in \bm{\Phi}(\psi),\\
            0; &\text{ otherwise },
        \end{cases}
    \end{aligned}
\end{equation}
where,
\begin{equation}
    \begin{aligned}
        &N_{\phi_\psi} = \text{ Number of phase connections to the network node,} \notag\\
        &S_\mathrm{base} = \text{ Base apparent power (kVA).}\notag
    \end{aligned}
\end{equation}\nomenclature[A]{\(N_{\phi_\psi}\)}{Number of phase connections of a DER $\psi$ to the network node;}\nomenclature[A]{\(S_\mathrm{base}\)}{Base apparent power in kVA;}
The per-phase reactive power injection by a DER at power factor $\theta(\psi)$ is given by
\begin{subequations}
    \begin{align}
    \label{q_psi}
    q^\phi(\psi)  = \eta(\psi) &\times p^\phi(\psi);\quad\text{ (p.u.) }\\
    \eta(\psi) = \sqrt{\frac{1}{[\theta(\psi)]^2} -1}&; \quad\theta(\psi) \in (0,1].
    \end{align}
\end{subequations}\nomenclature[A]{\(p^\phi(\psi),q^\phi(\psi)\)}{Per-phase real and reactive power injection of a DER \(\psi\) in p.u.;}\nomenclature[A]{\(\eta_\psi\)}{Ratio (unit-less) of reactive to active power of a DER \(\psi\);}
\vspace{-1em}
\section{Independent Distribution System Operator's T-DOPF for DER Aggregation}
\label{sec:model}
This section details the proposed IDSO's T-DOPF to perform value-based and grid-safe aggregation of the DERs participating within a bid-based TES design.
The objective function and constraints of the optimization model are described as follows:
\subsection{IDSO's T-DOPF Objective Function}
The objective of the IDSO is to select bids/offers for WPM participation, with the aim of efficiently serving DERs while maintaining distribution system reliability. The proposed IDSO model achieves this by first identifying the combination of bids/offers that maximizes the participants' net benefit without violating system constraints. To this end, the IDSO may fully clear, partially clear or reject bids and offers from DERs based on prevailing system conditions. This decision is facilitated by a variable characterised for each \(\psi\), $\alpha(\psi) \in [0,1],$\nomenclature[V]{\(\alpha(\psi)\)}{Decision variable to model clearing of a DER $\psi$;} as shown in (\ref{eq:power_sum}). The real and reactive power injection at any bus $i$ of the network can therefore be denoted as:
\begin{equation}
\label{eq:power_sum}
\begin{aligned}
         p^\phi_i &= \sum_{\psi \in \bm{\Psi}_i} [\alpha(\psi) p^\phi(\psi)] + p^{\phi, f}_i; \quad \forall \phi \in \bm{\Phi}, \quad \forall i \in \mathcal{N},\\
        q^\phi_i &= \sum_{\psi \in \bm{\Psi}_i} [\eta(\psi) \alpha(\psi)  p^\phi(\psi)] + q^{\phi, f}_i;\quad \forall \phi \in \bm{\Phi}, \quad \forall i \in \mathcal{N}.
\end{aligned}
\end{equation}
Here, $p^\phi(\psi)$ is the per-phase real power injection of $\psi^\text{th}$ DER in p.u., and $p^{\phi, f}_i \text{ and } q^{\phi, f}_i$\nomenclature[A]{$p^{\phi, f}_i,q^{\phi, f}_i$}{Fixed real and reactive power injections at a bus $i$ and phase $\phi$ in p.u.;} are the fixed real and reactive power injection at a distribution bus $i$ and phase $\phi$ respectively given in p.u. The DER active and reactive power injections at each phase and each node can be represented in vector notation by,
\begin{equation}
\label{eq:Vector_pq}
        \begin{aligned}
            \bm{p}_{1:N}^\Psi = \bm{A}_\Psi \bm{\alpha}_\Psi \bm{p}_\Psi; \quad
            \bm{q}_{1:N}^\Psi = \bm{A}_\Psi \bm{\eta}_\Psi \bm{\alpha}_\Psi \bm{p}_\Psi,
        \end{aligned}
\end{equation}\nomenclature[A]{\(\bm{A}_\Psi\)}{Node-DER incidence matrix;}\nomenclature[V]{\(\bm{\alpha}_\Psi\)}{Block diagonal matrix of decision variables, where each diagonal block is $\alpha(\psi)\cdot\bm{I}^3$ for each $\psi \in \Psi$;}\nomenclature[A]{$\bm{\eta}_\Psi$}{Block diagonal matrix of ratios of reactive to active power, where each diagonal block is $\eta(\psi)\cdot \bm{I}^3$ for each $\psi \in \Psi$;}\nomenclature[A]{$\bm{p}_\Psi$}{Column vector containing bid/offer volume for each $\psi \in \Psi$;}
where,
 $\bm{A}_\Psi \in \{\bm{I}^3, 0\}^{N \times |\Psi|}$ is the node-DER incidence matrix with the entries expressed as $\forall i \in \mathcal{N}, \forall \psi \in \Psi$:
    \begin{equation}
    \bm{A}_{i,\psi} = 
        \begin{cases}
            \bm{I}^3, \text{ if DER $\psi$ is connected to node $i$},\\
            0, \text{ otherwise}.
        \end{cases}
    \end{equation}
And,
\begin{equation*}
     \begin{aligned}
         \bm{\alpha}_\Psi \in \mathbb{R}^{3|\Psi| \times 3|\Psi|} =& \,\text{diag}((\alpha(\psi))_{\psi \in \Psi})\otimes \bm{I}^3, \\
         \bm{\eta}_\Psi \in \mathbb{R}^{3|\Psi| \times 3|\Psi|} =& \,\text{diag}((\eta(\psi))_{\psi \in \Psi})\otimes \bm{I}^3,\\
        \bm{p}_\Psi \in \mathbb{R}^{3|\Psi| \times 1} =&\, \left[ \left[ p^\phi(\psi) \right]_{\phi \in \bm{\Phi}}\right]_{\psi \in \Psi}.
     \end{aligned}
 \end{equation*}
Therefore, using the vector notation in \eqref{eq:Vector_pq}, (\ref{eq:power_sum}) can be equivalently represented as:
\begin{equation}
\label{eq:vectorpq1N}
        \begin{aligned}
            \bm{p}_{1:N} = \bm{p}_{1:N}^\Psi + \bm{p}_{1:N}^f; \quad
            \bm{q}_{1:N} = \bm{q}_{1:N}^\Psi + \bm{q}_{1:N}^f,
        \end{aligned}
\end{equation}\nomenclature[A]{\(\bm{p}_{1:N}^f, \bm{q}_{1:N}^f\)}{3-phase real and reactive power injection (p.u.) column vectors of fixed loads across all non-head buses in the distribution network;}\nomenclature[A]{\(\bm{p}_{1:N}^\Psi, \bm{q}_{1:N}^\Psi\)}{3-phase real and reactive power injection (p.u.) column vectors of DERs across all non-head buses in the distribution network;}
Here, $\bm{p}_{1:N}^\Psi, \bm{q}_{1:N}^\Psi \in \mathbb{R}^{3N \times 1}$ represents the column vectors of the DER active and reactive power injections at each phase at each node, respectively and $\bm{p}_{1:N}^f, \bm{q}_{1:N}^f \in \mathbb{R}^{3N \times 1}$ are column vectors that denote the fixed active and reactive power injections connected at each phase at each node, respectively. 
\par As stated earlier, the aggregation method under the proposed T-DOPF approves bids/offers with a higher net benefit or economic value, subject to constraints. Therefore, the IDSO's objective function for T-DOPF formulation to undertake grid-safe and value-based aggregation of bids and offers from DERs takes the following form:
    \begin{equation}
    \begin{aligned}
    \label{eq:IDSOOptimisation}
    \min_{\bm{\alpha}_\Psi, \bm{P}, \bm{Q}}  \bm{\gamma}_\Psi^T \bm{\alpha}_\Psi \bm{p}_\Psi S_\mathrm{base} \Delta t + C_\mathrm{IDSO}(\bm{P}).
    \end{aligned}
    \end{equation}\nomenclature[A]{\(C_\mathrm{IDSO}(\bm{P})\)}{Network cost of supplying power to the feeder in \textcent;}\nomenclature[A]{\(\Delta t\)}{Time period of operation in hours;}\nomenclature[A]{\(\bm{\gamma}_\Psi\)}{Column vector of bid/modified offer price in (\textcent/kWh) for each $\psi \in \Psi$;}\nomenclature[A]{\(M\)}{Big-M Number;}\nomenclature[A]{$\gamma(\psi)$}{Bid/modified offer price for a $\psi \in \Psi$;}
Here,
\begin{subequations}
    \begin{align*}
    &\gamma(\psi) = 
    \begin{cases}
        \pi(\psi); & \text{ if $\psi$ bids,}\\
        \pi(\psi) - \frac{M}{p(\psi)}; & \text{ if $\psi$ offers,}
    \end{cases}\\
    &\text{M is a \textit{big-M} Number,}\notag\\
    &\bm{\gamma}_\Psi \in \mathbb{R}^{3|\bm{\Psi}| \times 1} = [\gamma(\psi)]_{\psi \in \Psi} \otimes \bm{1}_3.
    \end{align*}\nomenclature[A]{$m$}{Marginal cost of providing distribution network services in \textcent/kWh;}
\end{subequations}
The objective function translates to a net-benefit maximization problem, where the IDSO selects bids/offers based on their economic value. The latter part of the objective function, $C_\mathrm{IDSO}(\bm{P}) = m \,\bm{1}^T_3\bm{c_0}^T\bm{P}$ (¢), models the cost of supplying power to the feeder; this may include the network's O\&M, reactive power support, and feeder reconfiguration costs, where, $m$ (\textcent/kWh), represents the marginal cost of providing distribution network services.
\begin{remark}
    Although the terms $\bm{A}_{i,\psi}$ and $\bm{\gamma}_{\Psi}$ may suggest that a DER is connected to all three phases, \textcolor{blue}{only} the actual phase connections will have $p^\phi(\psi)\neq0$ as modelled in \eqref{eq:PerPhase}. Thus, multiplying by $p^\phi(\psi)$ in the objective function reveals the actual phase connectivity of the DER.
\end{remark}
\vspace{-1.5em}
\subsection{IDSO's T-DOPF Technical Constraints}
The IDSO's objective function of the optimization formulation in (\ref{eq:IDSOOptimisation}) is subject to various technical constraints to operate the distribution system reliably. These constraints include:
\subsubsection{Power Balance Constraints} The active and reactive power balance constraints using \eqref{eq:MatrixLinDistFlowFinalpq} and \eqref{eq:vectorpq1N} are given by -
\begin{align}
    \label{eq:lambda_p}
    \bm{C}^T\bm{P} = \bm{p}_{1:N}^\Psi + \bm{p}_{1:N}^f, \\
    \label{eq:lambda_q} 
    \bm{C}^T\bm{Q} = \bm{q}_{1:N}^\Psi + \bm{q}_{1:N}^f.
\end{align}
\subsubsection{Nodal Voltage Magnitude Constraint}
The voltage magnitudes are constrained following the operating standard \cite{ANSI-C84.1-2020}.
\begin{equation}
\label{eq:v_alpha}
    \begin{aligned}
       \bm{v}_{min} \leq \bm{v}_{1:N} = \bm{I}^3_N\bm{v}_0 + 2\bm{C}^{-1} \left[ \bm{D_r} \bm{P} + \bm{D_x} \bm{Q} \right] \leq \bm{v}_{max}.
    \end{aligned}
\end{equation}
\subsubsection{Line Thermal Limits}
The feeder lines are constrained by their peak apparent power flow limit. 
Since these constraints are non-linear, this work employs the linearised constraints from \cite{LineAppr}, which uses a polygonal inner-approximation method to linearise the constraints. 
\begin{equation}
    \label{eq:LinThermalCap}
    \begin{aligned}
        \beta_e P^\phi_{(i,j)} &+ \delta_e Q^\phi_{(i,j)} 
        + \gamma_e S^\phi_{(i,j),\text{max}} \leq 0; \\
        &\quad \forall \phi \in \bm{\Phi},\ \forall (i,j) \in \mathcal{L},\ \forall e \in \mathcal{E}.
    \end{aligned}
\end{equation}
Here $\beta_e, \delta_e, \text{ and } \gamma_e$ are the coefficients of the linearised constraints where $e$ denotes an edge in the edge-set $\mathcal{E} = \{1,2,\dotsi,12\}$ of the polygonal approximation. Representing these using the column vector convention as:
\begin{align}
        \label{eq:LinThermalCapMatrix}
        &\beta_e \bm{P} + \delta_e \bm{Q} + \gamma_e \bm{S}_\mathrm{max} \leq \bm{0}_{3N}; \quad \forall e \in \mathcal{E}.
\end{align}\nomenclature[A]{$\beta_e, \delta_e, \gamma_e$}{Coefficients of the linearised constraints where $e$ denotes an edge in the edge-set $\mathcal{E}$;}\nomenclature[S]{$\mathcal{E}$}{Edge set of the polygon from the polygonal approximation of non-linear constraints;}
Here, $\bm{S}_\mathrm{max} \in \mathbb{R}^{3N \times 1} = [[S^\phi_{(i,j), \mathrm{max}}]_{\phi \in \bm{\Phi}}]_{(i,j)\in \mathcal{L}}$.\nomenclature[A]{$\bm{S}_\mathrm{max}$}{3-phase column vector of maximum apparent power flow limit for each line $(i,j) \in \mathcal{L}$;}\\
\subsubsection{IDSO's Substation Transformer Constraint}
IDSO's substation transformer limits the upstream/downstream transfer of power. This constraint can be represented in the linear form by using the same polygonal inner-approximation method as used before.
\begin{equation}
    \label{eq:LinSubstationLim}
    \begin{aligned}
       \beta_e \bm{p}_0 + \delta_e \bm{q}_0 + \gamma_e \bm{S}_{0,max} \leq \bm{0}_{3}; \quad \forall e \in \mathcal{E}.
    \end{aligned}
\end{equation}
Where, \(\bm{S}_{0,\mathrm{max}} \in \mathbb{R}^{3 \times 1}=[S^\phi_{0, \mathrm{max}}]_{\phi \in \bm{\Phi}}\) is the IDSO's substation transformer limit across the three phases.\nomenclature[A]{$\bm{S}_{0,\mathrm{max}}$}{3-phase column vector of IDSO's substation transformer limit across the three phases;}
\subsection{IDSO's Aggregation Problem}
The entire IDSO's T-DOPF mathematical formulation is given by:
\begin{equation}
\label{eq:FullOPF}
\begin{aligned}
  \min_{\bm{\alpha_\Psi}, \bm{P}, \bm{Q}}  \bm{\gamma}_\Psi^T \bm{\alpha}_\Psi \bm{p}_\Psi S_\mathrm{base} \Delta t + C_\mathrm{IDSO}(\bm{P})\\
\textit{subject to}:\quad 
\eqref{eq:lambda_p},
\eqref{eq:lambda_q},
\eqref{eq:v_alpha},
\eqref{eq:LinThermalCapMatrix},
\eqref{eq:LinSubstationLim}
\end{aligned}
\end{equation}\nomenclature[V]{\(\bm{\lambda}^p, \bm{\lambda}^q \)}{Row vectors of the dual multipliers (\textcent/p.u.) corresponding to the active and reactive power balance constraints;}\nomenclature[V]{\(\overline{\bm{\mu}}^v,\underline{\bm{\mu}}^v\)}{Row vectors of the dual multipliers (\textcent/p.u.) corresponding to the max and min voltage constraints;}
Let $\bm{\lambda}^p, \bm{\lambda}^q \in \mathbb{R}^{1 \times 3N}$ (¢/p.u.) denote the dual multipliers of the \textit{active and reactive power balance} constraints, respectively. The row vectors $\overline{\bm{\mu}}^v,\underline{\bm{\mu}}^v \in \mathbb{R}^{1 \times 3N}$ (¢/p.u.) denote the dual multipliers of the \textit{max and min voltage} constraint equation, respectively. Also, $\bm{\mu}^P(e) \in \mathbb{R}^{1 \times 3N} \text{ and } \bm{\mu}^{sub}(e) \in \mathbb{R}^{1 \times 3}$ (¢/p.u.) denote the dual multiplier of the \textit{line thermal limits} and \textit{substation transformer limit} constraints for an edge $e \in \mathcal{E}$. Since, for a given thermal line limit or substation transformer limit constraints violation, there can be more than one edge for which the constraints may become binding, we can group these edges into sets \(\mathcal{E}^P \text{ and } \mathcal{E}^{sub}\), to reflect for each constraint.
\begin{definition}
    \label{BOAProblem}
    The problem in \eqref{eq:FullOPF} is defined as \textit{IDSO's aggregation problem. The optimal solution $\bm{\alpha}^*_\Psi$ can be represented as \(\mathcal{A} \coloneq \{\alpha^*(\psi) = \bm{\alpha}^*_\Psi(3(\nu(\psi) -1)+1, 3(\nu(\psi)-1)+1) \,|\, \forall \psi \in \bm{\Psi} \},\), where \(\nu \colon \bm{\Psi} \to \{1,2,\ldots,|\bm{\Psi}|\},\) such that, \(\nu(\psi_1) = \nu(\psi_2) \implies \psi_1 = \psi_2,\quad \forall\, \psi_1,\psi_2 \in \bm{\Psi}.\)} 
\end{definition}
\begin{definition}
\label{NQP}
    \textit{For IDSO's aggregation problem in (\ref{eq:FullOPF}), the vector of optimal dual multipliers, $\bm{\lambda}^{p^{*}}, \bm{\lambda}^{q^{*}} \in \mathbb{R}^{1 \times 3N}$ are named the Active-Nodal Qualification Prices (A-NQPs) and Reactive-Nodal Qualification Prices (R-NQPs) in (\textcent/p.u.), respectively.}
\end{definition}
\nomenclature[S]{\(\mathcal{A}\)}{Optimal solution set of \(\alpha^*(\psi)\,\, \forall \,\,\psi\in\bm{\Psi}\) ;}
\nomenclature[V]{\(\bm{\mu}^P(e)\)}{Row vector of thermal line limit dual multipliers (\textcent/p.u.) for a particular \(e \in \mathcal{E}\);}\nomenclature[V]{\(\bm{\mu}^{sub}(e)\)}{Row vector of IDSO's substation limit dual multipliers (\textcent/p.u.) for a particular \(e \in \mathcal{E}\);}
The above T-DOPF optimization formulation will be used in the proposed bids/offers aggregation method, as described in the next section.
\vspace{-0.9em}
\section{Methodology}\label{sec:methodology}
This section outlines the proposed methodology for DER aggregation and WPM participation. It builds upon the T-DOPF in \eqref{eq:FullOPF} to qualify bids and offers for participation in WPM while accounting for the T-D interactions inherent in an ITD system. The study first addresses the scenario where the DERs only have bidding capabilities, and then it addresses the complementary scenario where the DERs only have offering capabilities. Finally, the general case where DERs can have both bidding and offering capabilities is addressed. As explained in the next subsections, the proposed bid/offer prequalification-cum-aggregation method, together with the IDSO's WPM participation and retail price determination methods, accounts for the two-way T-D interactions between IDSO and ITSO (exchange of bid/offer functions and LMPs) without making assumptions regarding either the LMP or the total demand.
\vspace{-1em}
\subsection{Case A: DERs have only bidding capabilities}\label{case:bid}
\subsubsection{Bid Prequalification-cum-Aggregation Method}
IDSO receives bids from the participating DERs and performs the proposed grid-safe and value-based aggregation. For this, the IDSO executes the T-DOPF in \eqref{eq:FullOPF} and arranges the results of the T-DOPF as given by --
\begin{equation}
   \begin{aligned}
    \bm{\alpha} =& \{ \alpha(\psi) = \alpha^*(\psi) \in \mathcal{A} \;\Bigm|\; \psi \in \bm{\Psi}^- \}.
    \end{aligned} 
\end{equation}
The A-NQPs and R-NQPs, as defined in Def.~\ref{NQP}, determine which bid qualifies for participation in WPM. The important results that these NQPs signify are stated in Proposition~\ref{prop1bid}.
\begin{prop}
\label{prop1bid}
    \textit{The NQPs satisfy the following relations:}
\begin{enumerate}
    \item For qualified bids, \(\forall i \in\mathcal{N},\,\forall\psi \in \bm{\Psi}^-_i| \alpha(\psi) \neq 0\), \\$\displaystyle\pi^\mathrm{QP}(\psi) \coloneq \frac{\sum_{\phi \in \bm{\Phi}(\psi)}(-\lambda_{i,\phi}^p  - \eta(\psi) \lambda_{i,\phi}^q)}{N_{\phi_\psi} S_\mathrm{base}\Delta t} \leq \pi(\psi);$ \label{Diffpbid}\\
     \item $\displaystyle \bm{C}\bm{\lambda}^{p^{*T}} = \nabla_{\bm{P}}C_\mathrm{IDSO}(\bm{P^*}) + 2\bm{D}_r^TC^{-T}[(\overline{\bm{\mu}}^{{v}^*})^T - (\underline{\bm{\mu}}^{{v}^*})^T] \\+ \sum_{e \in \mathcal{E}^P}[\beta_e(\bm{\mu}^{{P}^*}(e))^T] + \sum_{e \in \mathcal{E}^{sub}}[\beta_e\bm{C}_0 (\bm{\mu}^{{sub}^*}(e))^T];$ \label{DiffPbidprop}\\
    \item $\displaystyle \bm{C}\bm{\lambda}^{q^{*T}} = 2\bm{D}_x^TC^{-T}[(\overline{\bm{\mu}}^{{v}^*})^T - (\underline{\bm{\mu}}^{{v}^*})^T] \\+ \sum_{e \in \mathcal{E}^P}[\delta_e(\bm{\mu}^{{P}^*}(e))^T] + \sum_{e \in \mathcal{E}^{sub}}[\delta_e\bm{C}_0 (\bm{\mu}^{{sub}^*}(e))^T].$ \label{DiffQbidprop}\\
\end{enumerate}
\end{prop}\nomenclature[A]{\(\pi^\mathrm{QP}(\psi)\)}{Qualification price in (\textcent/kWh) for a $\psi \in \Psi$;}\nomenclature[A]{\(\nabla_{\bm{P}}C_\mathrm{IDSO}(\bm{P})\)}{Marginal network cost of supplying power to the feeder in \textcent/kWh;}
Prop.~\ref{prop1bid}.\ref{Diffpbid} provides the qualification or cut-off prices ($\pi^\mathrm{QP}(\psi)$) that establish limits on the bid prices for each $\psi \in \Psi^-$. Prop.~\ref{prop1bid}.\ref{DiffPbidprop} and Prop.~\ref{prop1bid}.\ref{DiffQbidprop} provide the relation of A-NQPs and R-NQPs with IDSO's marginal cost of providing network services and optimal dual multipliers of the network's inequality constraints. The propositions indicate how the additional price markup over and above the marginal cost of network operation comes into effect due to the activation of the inequality constraints. The proof of this proposition is provided in Appendix~\ref{Append:Prop1}.
\subsubsection{IDSO's WPM Participation} After qualification of bids, IDSO needs to participate in the WPM in order to meet their power needs and provide services to the grid. Since bids reflect the customer's maximum willingness to pay for a specified quantity of power---and thus set an upper bound on acceptable prices---the IDSO must ensure that any DER cleared in the WPM at LMP can cover the incurred network costs. As the IDSO cannot observe the LMP in advance, it must submit only those bids that can adequately cover these costs. To achieve this, IDSO constructs its bids for every \(\psi \in \bm{\Psi}^-\) satisfying \(\alpha(\psi) \neq 0\) as
\begin{align}
    \label{eq:IDSO'sBidPrice}
    \pi^\mathrm{IDSO}(\psi) &= \pi(\psi) - m,\\
    \label{eq:IDSO'sBidQuantity}
    p^\mathrm{IDSO}(\psi) &= \alpha(\psi) p(\psi).
\end{align}
By employing the above-constructed bid, the last cleared bid price from the DER is guaranteed to be at least $LMP +\,m$, ensuring cleared bids can recover network costs. After the WPM clears, the DERs which are cleared from the market are given by:
\begin{align}
    \mathcal{C}_{\bm{\Psi}}^-
   = \Bigl\{\,
       \psi \in \bm{\Psi^{-}}
       \,\Bigm|\,
       \alpha(\psi) \neq 0
       \;\wedge\;
       \pi(\psi) \ge LMP + m
     \Bigr\}.
\end{align}\nomenclature[S]{$\mathcal{C}_{\bm{\Psi}}^-$}{Set containing qualified DERs with bids cleared in the WPM;}
\subsubsection{Determination of Retail Prices}
For any TES design within a distribution system, retail price determination is a key aspect. The retail prices should be structured such that only those bids qualified by the IDSO and cleared in the WPM are accepted for power consumption. For the rejected bids, the retail price signals should indicate the price that would have ensured qualification for WPM participation and subsequent clearing. Meanwhile, all accepted and cleared DERs will receive a uniform price. The proposed ``differential pricing" to achieve the aforementioned objective is outlined as follows:
\par Let \(\overline{\bm{\Psi}^-}:= \{\psi \in \bm{\Psi}^- \;|\;\alpha(\psi) = 0\}\) denote the set containing DERs with unqualified bids, then the retail price signals to be assigned to the DERs after participation/clearance in WPM are given by,\nomenclature[S]{$\overline{\bm{\Psi}^-}$}{Set containing DERs with unqualified bids;}
\begin{definition}
    \label{retailbid}
    For a DER \(\psi \in \bm{\Psi}^-\) participating in a TES design for an unbalanced distribution network, the retail signals that IDSO projects subject to clearance in WPM are given by
    \begin{align}
        \pi^\mathrm{ret}(\psi) :=\begin{cases}
            \max(LMP +m, \, \pi^\mathrm{QP}(\psi)), &\forall\psi \in \overline{\bm{\Psi}^-},\\
            LMP + m, &\forall \psi \in \bm{\Psi}^-\setminus\overline{\bm{\Psi}^-}.
        \end{cases}
        \end{align}
    \begin{align}
        p^\mathrm{ret}(\psi) := \begin{cases}
            0, &\forall\psi \in \overline{\bm{\Psi}^-},\\
            \alpha(\psi) p(\psi), &\forall \psi \in \bm{\Psi}^-\setminus\overline{\bm{\Psi}^-}.
        \end{cases}
    \end{align}
\end{definition}
\vspace{-1em}
\subsection{Case B: DERs have only offering capabilities} \label{case:offer}
\subsubsection{Offer Prequalification-cum-Aggregation Method}
IDSO executes \eqref{eq:FullOPF} to solve its aggregation problem. The results are then categorised as:
\begin{align}
    \bm{\alpha} =& \{ \alpha(\psi) = \alpha^*(\psi) \in \mathcal{A} \;\Bigm|\; \psi \in \bm{\Psi}^+ \}.
\end{align}
The NQPs are again used to determine the cut-off prices to qualify offers for participation in WPM, as stated in Proposition~\ref{offerprop}.
\begin{prop}
    \label{offerprop}
    \textit{The NQPs satisfy the following relation:}
    \begin{enumerate}
        \item For qualified offers,\(\forall i \in\mathcal{N},\,\forall\psi \in \bm{\Psi}^+_i| \alpha(\psi) \neq 0,\)\\ $\pi^\mathrm{QP}(\psi) \coloneq \frac{\sum_{\phi \in \bm{\Phi}(\psi)}(-\lambda_{i,\phi}^p - \eta(\psi) \lambda_{i,\phi}^q)}{N_{\phi_\psi} S_\mathrm{base}\Delta t} + \frac{M}{p(\psi)} \geq \pi(\psi);$ \label{Diffpoffer}
    \end{enumerate}
\end{prop}
The proof of this proposition is in Appendix~\ref{Append:prop2}.
\begin{remark}
    Propositions~\ref{prop1bid}.\ref{DiffPbidprop} and \ref{prop1bid}.\ref{DiffQbidprop} remain valid for the case where participating DERs have only offering capabilities; the proofs are identical and therefore omitted.
\end{remark}
\subsubsection{IDSO's WPM Participation}
After the offers from DERs have been qualified, the IDSO participates in the WPM with the aggregated quantities. The approach used for offers in Case B is essentially the counterpart of the mechanism for bids as described for Case A. Since, in this case, the offers reflect the customer's minimum acceptance price---establishing a lower bound---the IDSO must ensure that any DER cleared in the WPM can adequately cover the incurred network costs. For this, it constructs its offer for every \(\psi \in \bm{\Psi}^+\), satisfying \(\alpha(\psi) \neq 0\) as,
\begin{align}
    \label{eq:IDSO'sOfferPrice}
    \pi^\mathrm{IDSO}(\psi) &= \pi(\psi) + m,\\
    \label{eq:IDSO'sOfferQuantity}
    p^\mathrm{IDSO}({\psi}) &= \alpha(\psi) p(\psi).
\end{align}
Through this, IDSO ensures that any compensation provided to DER for power production, the last cleared offer price is guaranteed to be at most $LMP - m$, ensuring recovery of network costs. The cleared DERs from the market are expressed as:
\begin{align}
    \mathcal{C}_{\bm{\Psi}}^+
   = \Bigl\{\,
       \psi \in \bm{\Psi^{+}}
       \;\Bigm|\;
       \alpha(\psi) \neq 0
       \;\wedge\;
       \pi(\psi) \le LMP - m
     \Bigr\}.
\end{align}\nomenclature[S]{$\mathcal{C}_{\bm{\Psi}}^+$}{Set containing qualified DERs with offers cleared in the WPM;}
\subsubsection{Determination of Retail Prices} Similar to case A, the retail price signal developed is based on ``differential pricing", where all cleared and accepted DERs in the WPM receive a uniform price, while the unqualified DERs are given a price that could have ensured qualification in WPM and subsequent clearing. Let \(\overline{\bm{\Psi}^+}:=\{\psi \in \bm{\Psi}^+\;|\;\alpha(\psi) = 0\}\), then retail prices signals assigned to DERs are given by:\nomenclature[S]{$\overline{\bm{\Psi}^+}$}{Set containing DERs with unqualified offers;}
\begin{definition}
    \label{def:retailoffer}
    For a DER $\psi \in \bm{\Psi}^+$ participating in a TES design for an unbalanced distribution network, the retail signals the IDSO projects subject to clearance in WPM, are given by
\begin{align}
    \pi^\mathrm{ret}(\psi) :=
            \begin{cases}
            \min \, (LMP - m, \,\pi^\mathrm{QP}(\psi)), &\forall\,\psi \in \overline{\bm{\Psi}^+},\\ 
            LMP - m, &\forall\,\psi \in \bm{\Psi}^+ \setminus \overline{\bm{\Psi}^+}.
            \end{cases}
\end{align}
\begin{align}
    p^\mathrm{ret}(\psi) :=
        \begin{cases}
            0, &\forall \psi \in \overline{\bm{\Psi}^+},\\
            \alpha(\psi) p(\psi), &\forall \psi \in \bm{\Psi}^+ \setminus \overline{\bm{\Psi}^+}.
        \end{cases}
\end{align}
\end{definition}
\subsection{Case C: DERs have both bidding and offering capabilities}\label{sec:CaseC}
\subsubsection{Bid/Offer Prequalification-cum-Aggregation Method}\label{sec:CaseC_1}
\begin{figure}[!htbp]
    \centering
    \includegraphics[width=1\linewidth]{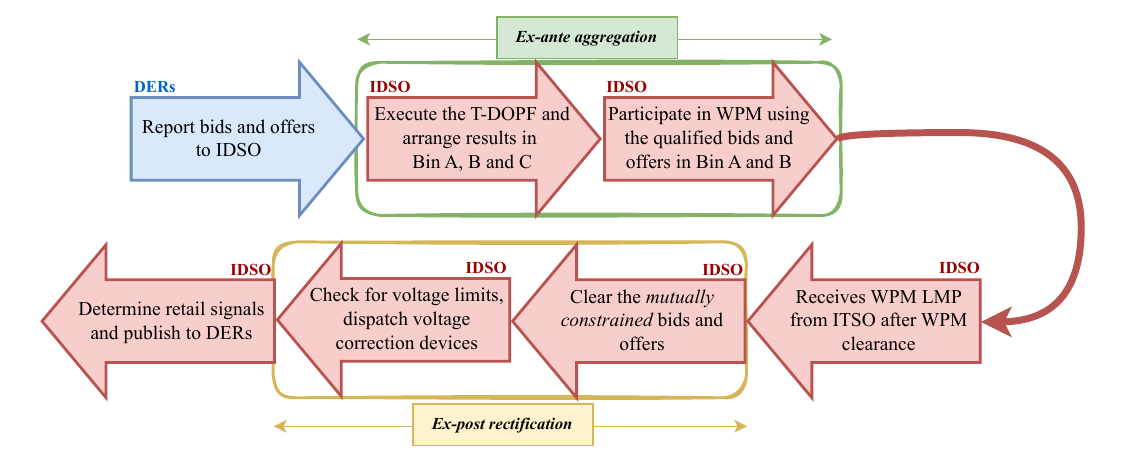}
    \caption{Flowchart for the proposed methodology to handle Case C.}
    \label{fig:flowchart}
\end{figure}

Similar to previous cases, IDSO solves its aggregation problem in \eqref{eq:FullOPF} and arranges the results in a Bin C as:
\begin{align}
\label{eq:alpha_c}
        \bm{\alpha_{C}} =&\{ \alpha_{C}(\psi) = \alpha^*(\psi) \in \mathcal{A} \;|\; \psi \in \bm{\Psi}\}.
\end{align} 
However, when DERs within the distribution system are submitting both bids and offers, the scenario becomes challenging for the IDSO---because the clearance of these bids and offers is interdependent on each other---within the IDSO's aggregation problem in \eqref{eq:FullOPF}. If these interdependent bids and offers are not cleared simultaneously in the WPM, they may cause contractual and network violations. Hence, proper \textit{ex-ante aggregation} and \textit{ex-post rectification} techniques need to be employed. The Fig.~\ref{fig:flowchart} depicts the methodology for this case. First, the \textit{ex-ante aggregation} technique is discussed, and the \textit{ex-post rectification} technique is presented subsequently, following the description of the IDSO's WPM participation.
\subsubsection*{Ex-ante Aggregation} As discussed before, the IDSO first determines the interdependent bids and offers that are mutually contingent on each other's clearance in WPM. For this, in addition to \eqref{eq:alpha_c}, it creates: Bin A, solving T-DOPF only for bids; Bin B, solving T-DOPF only for offers, similar to Case A and Case B, as shown below: 
\begin{equation}
   \begin{aligned}
    \bm{\alpha_A} =& \{ \alpha_A(\psi) = \alpha^*(\psi) \in \mathcal{A} \;|\; \psi \in \bm{\Psi}^- \},\\
    \bm{\alpha_{B}} =& \{ \alpha_{B}(\psi) = \alpha^*(\psi) \in \mathcal{A} \;|\; \psi \in \bm{\Psi}^+\}.
\end{aligned} 
\end{equation}\nomenclature[S]{\(\bm{\alpha}_A\)}{Set containing \(\alpha^*_\psi \) that are present in Bin A;}\nomenclature[S]{\(\bm{\alpha}_{B}\)}{Set containing  \(\alpha^*_\psi\) that are present in Bin B;}\nomenclature[S]{\(\bm{\alpha}_{C}\)}{Set containing \(\alpha^*_\psi\) that are present in Bin C;}
Then IDSO proceeds as follows to create the set of mutually constrained DERs:
\begin{align}
    \bm{\Psi}_{MC} = \{\psi \in \bm{\Psi} \,|\,\alpha_{MC}(\psi) \neq 0\}.
\end{align}
Where \(\alpha_{MC} (\psi)\) is a member of set,
\begin{equation}
\label{eq:alphamc}
\begin{aligned}
\bm{\alpha}_{MC}
&= \Bigl\{
   \alpha_{MC}(\psi) = \alpha_{C}(\psi) : \psi \in \bm{\Psi}
   \;\Big|\; \\
&\quad
   \begin{cases}
     \alpha_{C}(\psi) \neq \alpha_{B}(\psi);
     &\text{if } \psi \in \bm{\Psi}^+,\\
     \alpha_{C}(\psi) \neq \alpha_{A}(\psi);
     &\text{if } \psi \in \bm{\Psi}^-
   \end{cases}
\Bigr\}.
\end{aligned}
\end{equation}\nomenclature[S]{\(\bm{\alpha}_{MC}\)}{Set containing \(\alpha^*_\psi\,\, \forall\,\,\psi\in \bm{\Psi}_{MC}\);}\nomenclature[S]{\(\bm{\Psi}_{MC}\)}{Set of all mutually contingent DERs;}
The NQPs providing the cut-off or qualification prices are expressed in the proposition below:
\begin{prop}
    \label{eitherprop}
    \textit{The NQPs satisfy the following relations:}
\begin{enumerate}
    \item For qualified bids, \(\forall i \in\mathcal{N},\,\forall\psi \in \bm{\Psi}^-_i| \alpha_A(\psi) \neq 0\), \\$\pi^\mathrm{QP}(\psi) \coloneq \frac{\sum_{\phi \in \bm{\Phi}(\psi)}(-\lambda_{A,i,\phi}^p  - \eta(\psi) \lambda_{A,i,\phi}^q)}{N_{\phi_\psi} S_\mathrm{base}\Delta t} \leq \pi(\psi);$ \label{Diffpeitherbid}\\
    \item For qualified offers,\(\forall i \in\mathcal{N},\,\forall\psi \in \bm{\Psi}^+_i| \alpha_{B}(\psi) \neq 0,\)\\ $\pi^\mathrm{QP}(\psi) \coloneq\frac{\sum_{\phi \in \bm{\Phi}(\psi)}(-\lambda_{B,i,\phi}^p - \eta(\psi) \lambda_{B,i,\phi}^q)}{N_{\phi_\psi} S_\mathrm{base}\Delta t} + \frac{M}{p(\psi)} \geq \pi(\psi);$ \label{Diffpeitheroffer}
\end{enumerate}
\end{prop}
This proposition gives the qualification prices for bids and offers in Bin A and Bin B. The proof of this proposition is in Appendix~\ref{Append:prop3}. 
\begin{remark}
    Propositions~\ref{prop1bid}.\ref{DiffPbidprop} and \ref{prop1bid}.\ref{DiffQbidprop}, remain valid for the case where participating DERs have both bidding and offering capabilities; the proofs are identical and therefore omitted.
\end{remark}
\subsubsection{IDSO's WPM Participation} \label{sec:WPMNW}
Analogous to the cases A and B, the IDSO only uses the qualified bids and offers categorised in Bin A and Bin B, respectively, to participate in the WPM. It does so by formulating its own bids and offers by factoring in the network costs, as shown in \eqref{eq:IDSO'sBidPrice}, \eqref{eq:IDSO'sBidQuantity}, \eqref{eq:IDSO'sOfferPrice}, and \eqref{eq:IDSO'sOfferQuantity} as
\begin{align}
    \label{eq:IDSO'sBidOfferPrice}
    \pi^\mathrm{IDSO}({\psi}) = 
    \begin{cases}
        \pi(\psi) - m, & \psi \in \bm{\Psi}^- \text{ and } \alpha_{A}(\psi) \neq 0, \\
        \pi(\psi) + m, & \psi \in \bm{\Psi}^+ \text{ and } \alpha_{B}(\psi) \neq 0.
    \end{cases} 
\end{align}
\begin{align}
     \label{eq:IDSO'sBidOfferQuantity}
    p^\mathrm{IDSO}({\psi}) = \alpha(\psi) p(\psi); \quad &\psi \in \bm{\Psi} \text{ and } \\&(\alpha_A(\psi) \neq 0 \text{ or } \alpha_{B}(\psi) \neq 0).\notag
\end{align}\nomenclature[A]{\(\pi^\mathrm{IDSO}({\psi}), p^\mathrm{IDSO}({\psi})\)}{IDSO's modified bid/offer price (\textcent/kWh) and volume (kW) pair for a qualified DER $\psi$ after accounting for network costs;}
Similar to previous cases, the cleared DERs are given by:
\begin{equation}
    \begin{aligned}
    \mathcal{C}_{\bm{\Psi}}^-
   = \Bigl\{\,
       \psi \in \bm{\Psi^{-}}
       \;\Bigm|\;
       \alpha_A(\psi) \neq 0
       \;\wedge\;
       \pi(\psi) \ge LMP + m
     \Bigr\}.\\
       \mathcal{C}_{\bm{\Psi}}^+
   = \Bigl\{\,
       \psi \in \bm{\Psi^{+}}
       \;\Bigm|\;
       \alpha_B(\psi) \neq 0
       \;\wedge\;
       \pi(\psi) \le LMP - m
     \Bigr\}.
\end{aligned}
\end{equation}
Since Bin C contains interdependent bids and offers, they are purposefully retained and are reconsidered during ex-post WPM operations. The corresponding procedure is described below.
\subsubsection*{Ex-post Rectification}
\label{sec:MCBidsOffers}
As previously discussed, the IDSO is aware of the mutually constrained bids and offers as specified in \eqref{eq:alphamc}. Furthermore, since the IDSO obtains the LMP at the linkage bus after the WPM clears, it can schedule mutually contingent bids and offers additionally while ensuring that the net interchange at the T-D linkage does not exceed the contracted amount to avoid any penalty for deviations. 
For this, the IDSO first determines the subset of mutually constrained DERs that can cover the network and energy costs given by
\begin{align}
    \bm{\tilde{\Psi}}_{MC}=\{\psi \in \bm{\Psi}_{MC}\; | \;\tilde{\alpha}_{MC}(\psi)\neq0\},
\end{align}
where,
\begin{align}
&\bm{\tilde{\alpha}}_{MC}
= \Bigl\{
   \tilde{\alpha}_{MC}(\psi) =  \alpha_{MC}(\psi) \cdot \mathbbm{1}_{\{f(\pi(\psi)) \le 0\}} : \psi \in \bm{\Psi}_{MC}
    \;|\;  \notag \\
&\begin{cases}
      f(\pi(\psi)): \pi(\psi) - LMP
                   + m;
       & \text{if } \psi \in \bm{\Psi}^+, \\[4pt]
      f(\pi(\psi)): -\pi(\psi) + LMP
                   + m;
       & \text{if } \psi \in \bm{\Psi}^-.
    \end{cases}
\Bigr\},
\end{align}
here \(\mathbbm{1}_{\{f(\pi(\psi)) \le 0\}}\) is an indicator which equals to $1$ if $f(\pi(\psi)) \le 0$ and 0 otherwise.\nomenclature[S]{\(\bm{\tilde{\alpha}}_{MC}\)}{Set of \(\tilde{\alpha}_{MC}(\psi)\) corresponding to mutually contingent DER, $\psi \in \bm{\Psi}_{MC}$ that obey network and energy costs;}\nomenclature[A]{$LMP$}{LMP observed by the IDSO at WPM (\textcent/kWh);}\nomenclature[S]{\(\bm{\tilde{\Psi}}_{MC}\)}{Subset of the set of mutually contingent DERs \(\bm{\Psi}_{MC}\) that obey network and energy costs.}
Subsequently, the IDSO re-runs the T-DOPF  in \eqref{eq:FullOPF} for DERs in \(\tilde{\bm{\Psi}}_{MC}\) with two additional constraints--
\begin{align}
\label{eq:fixalpha}
    \alpha(\psi) = \begin{cases}
        \alpha_A(\psi), \quad &\forall \psi \in \mathcal{C}_{\bm{\Psi}}^{-}, \\
        \alpha_B(\psi), \quad &\forall \psi \in \mathcal{C}_{\bm{\Psi}}^{+}, \\
        0, \quad &\forall \psi \in \bm{\Psi}\setminus\{\mathcal{C}_{\bm{\Psi}}^{-}\cup\mathcal{C}_{\bm{\Psi}}^{+}\cup \tilde{\bm{\Psi}}_{MC}\}
    \end{cases} 
\end{align}
\begin{align}
\label{eq:netinter}
    \sum_{\psi \in \tilde{\Psi}_{MC}}\alpha(\psi) p(\psi) = 0.
\end{align}
The constraint in \eqref{eq:fixalpha} fixes the variables corresponding to cleared bids and offers as constant injections, while simultaneously setting the variables associated with unqualified bids and offers to zero injections. Eqn~\eqref{eq:netinter} constrains the net additional volume of selected mutually constrained bids and offers to remain zero, ensuring that the results of WPM are not violated.
The results are then given by \(\bm{\hat{\alpha}}_{MC} = \{\hat{\alpha}_{MC}(\psi)=\alpha^*(\psi) \in \mathcal{A}\:|\;\psi \in \tilde{\Psi}_{MC}\}\) such that cleared mutually constrained DERs (\(\mathcal{C}^{MC}_{\bm{\Psi}} = \{\psi \in \bm{\tilde{\Psi}}_{MC}\;|\;\hat{\alpha}_{MC}(\psi)\neq0\}\)) neither affect the network constraints nor violate the scheduled net interchange.\nomenclature[S]{\(\bm{\hat{\alpha}}_{MC}\)}{Set containing optimal \(\alpha^*(\psi)\) for $\psi \in \bm{\tilde{\Psi}}_{MC}$ after running T-DOPF for mutually contingent DERs that obey network and energy costs ($\tilde{\bm{\Psi}}_{MC}$);}
Hence, the  cleared DERs in this case are given by:
\begin{align}
\mathcal{C}_{\bm{\Psi}} = \mathcal{C}_{\bm{\Psi}}^{+}\cup\mathcal{C}_{\bm{\Psi}}^{-}\cup\mathcal{C}_{\bm{\Psi}}^{MC}.
\end{align}\nomenclature[S]{$\mathcal{C}_{\bm{\Psi}}^{MC}$}{Set containing cleared mutually contingent DERs;}
\begin{remark}
    The proposed design employs a Three-Phase LinDistFlow formulation, a widely employed linear approximation of the distribution power flow. Once the cleared bids and offers are obtained, the IDSO needs to carry out routine network violation checks to enable seamless real-time operation. If any violations are found, IDSO may exercise the volt-var control of DERs compliant with the IEEE 1547 standard \cite{IEEE1547-2018} for reliable network operations. Although the DERs need to be compensated for those services, such a pricing mechanism is beyond the scope of this study and will be addressed in future work.
\end{remark}
\subsubsection{Determination of Retail Prices}
\label{sec:RetailDesign}
Retail prices follow the same principle of ``differential" pricing as the previous two cases to achieve the objective of prices that are indicative of the true value of procuring or offering services within the distribution system.
Let \(\overline{\bm{\Psi}}_A := \{\psi \in \bm{\Psi}^-\;|\;(\alpha_A(\psi)= 0) \wedge (\tilde{\alpha}_{MC}(\psi) = 0 \,\vee\, \hat{\alpha}_{MC}(\psi)=0)\},\) denote the set containing DERs with unqualified bids, \( \overline{\bm{\Psi}}_{B} \coloneq \{\psi \in \bm{\Psi}^+\;|\;(\alpha_{B}(\psi)= 0) \wedge (\tilde{\alpha}_{MC}(\psi) = 0 \,\vee\, \hat{\alpha}_{MC}(\psi)=0)\}, \) be the set containing DERs with unqualified offers. Also, let the sets containing all cleared bids be denoted by \(\bm{\Psi}_A \coloneq \{ \psi \in \bm{\Psi}^-\;|\;\alpha_A(\psi) \neq 0 \,\vee\, \hat{\alpha}_{MC}(\psi) \neq 0\}\), and all cleared offers by \(\bm{\Psi}_{B} = \{ \psi \in \bm{\Psi}^+\;|\;\alpha_{B}(\psi) \neq 0 \,\vee\, \hat{\alpha}_{MC}(\psi) \neq 0 \}\). The following definition sets up the retail signals to be sent to participating DERs.
\begin{definition}
    \label{def:retail}
    For a DER $\psi \in \bm{\Psi}$ participating in a bid-based TES design for an unbalanced distribution network, the retail signals the IDSO projects for DERs, subject to WPM clearing, are given by:
\begin{equation}
    \begin{aligned}
    \label{eq:retail} 
            \pi^\mathrm{ret}(\psi) &\coloneq
            \begin{cases}
           \max \, (LMP + m,\ \ \pi^\mathrm{QP}(\psi)),&\forall\,\psi \in \overline{\bm{\Psi}}_A,\\
            \min \, (LMP - m,\ \ \pi^\mathrm{QP}(\psi)),&\forall\,\psi \in \overline{\bm{\Psi}}_B,\\ 
            LMP + m, &\forall\,\psi \in \bm{\Psi}_A,\\
            LMP - m, &\forall\,\psi \in \bm{\Psi}_B.\\
            \end{cases}
    \end{aligned}
\end{equation}
\begin{equation}
    \begin{aligned}
        p^\mathrm{ret}(\psi) \coloneq 
        \begin{cases}
            0, \quad &\forall \psi \in \overline{\bm{\Psi}}_A \cup \overline{\bm{\Psi}}_B,\\
            \alpha(\psi) p(\psi), \quad &\forall \psi \in \bm{\Psi}_A \cup \bm{\Psi}_B.
        \end{cases}
    \end{aligned}
\end{equation}
\end{definition}\nomenclature[A]{\(\pi^\mathrm{ret}(\psi), p^\mathrm{ret}(\psi)\)}{Retail price (\textcent/kWh) and quantity (kW) sent to DER $\psi \in \bm{\Psi}$}\nomenclature[S]{\(\overline{\bm{\Psi}}_A\)}{Set containing DERs with unqualified bids;}\nomenclature[S]{\(\overline{\bm{\Psi}}_{B}\)}{Set containing DERs with unqualified offers;}\nomenclature[S]{\(\bm{\Psi}_{B}\)}{Set containing DERs with qualified offers;}\nomenclature[S]{\(\bm{\Psi}_{A}\)}{Set containing DERs with qualified bids;}
The proposed method, as described in this section and demonstrated in the following section using sample test cases, accounts for the T-D interactions of realistic ITD systems, incorporates network costs, and provides value-based retail prices.
\section{Results and Discussion} \label{sec:Results}
The proposed methods are tested using a modified IEEE 123-bus radial network populated by DERs randomly across different phases and buses. The results advocate the efficacy of the proposed design, as presented in the next subsections.
\subsection{Test System and Preliminaries} \label{sec:TestSP}
\begin{figure}[!htbp]
    \centering
    \includegraphics[width=0.9\linewidth, height=3.8cm]{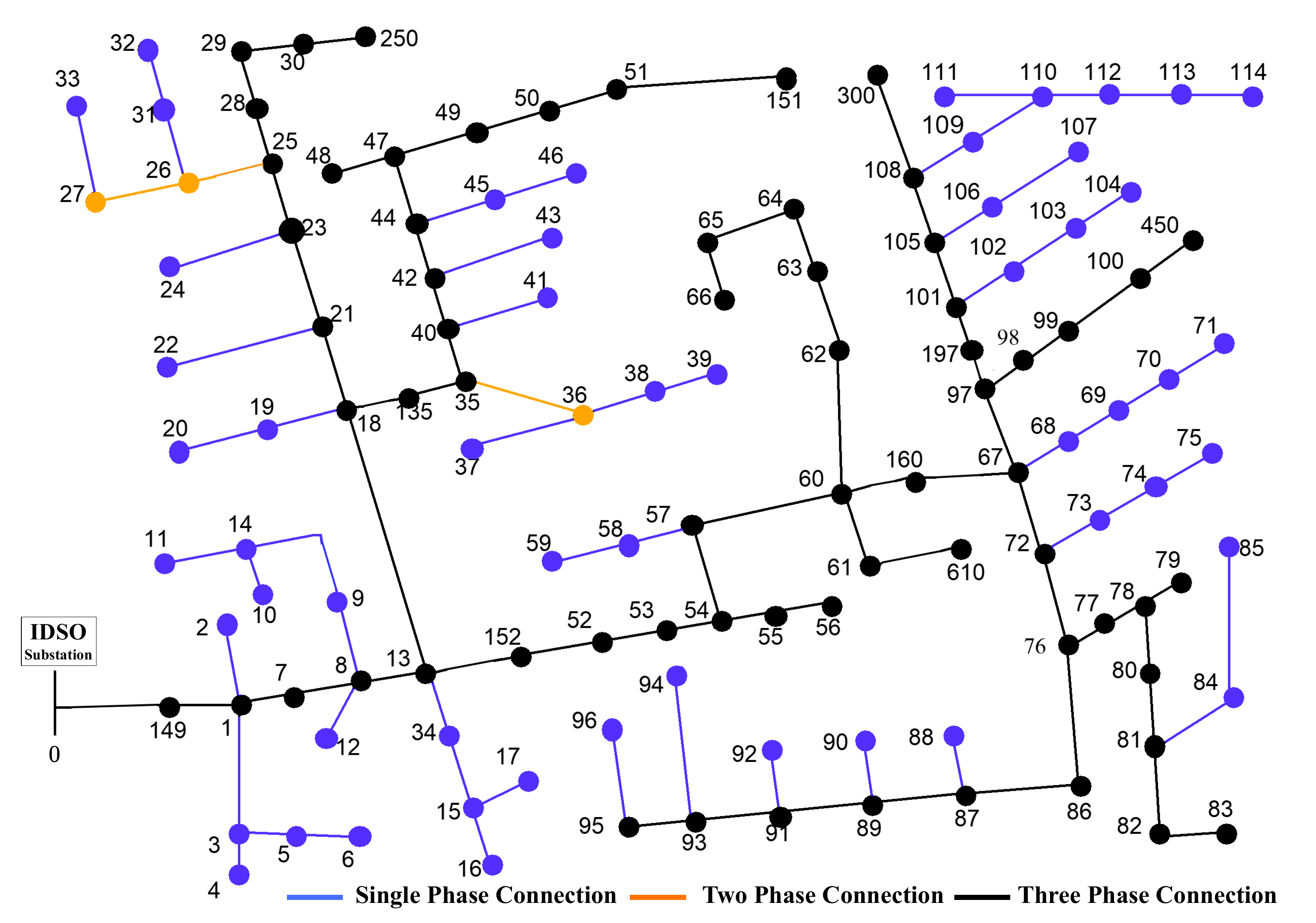}
    \caption{Modified IEEE 123 bus radial distribution network.}
    \label{fig:IEEE123}
\end{figure}
The standard IEEE 123-bus radial network \cite{Kersting2007} has been modified as shown in Fig. \ref{fig:IEEE123}. DERs are integrated randomly at different buses and phase connections. 
\begin{figure*}
    \centering
    \includegraphics[width=\linewidth]{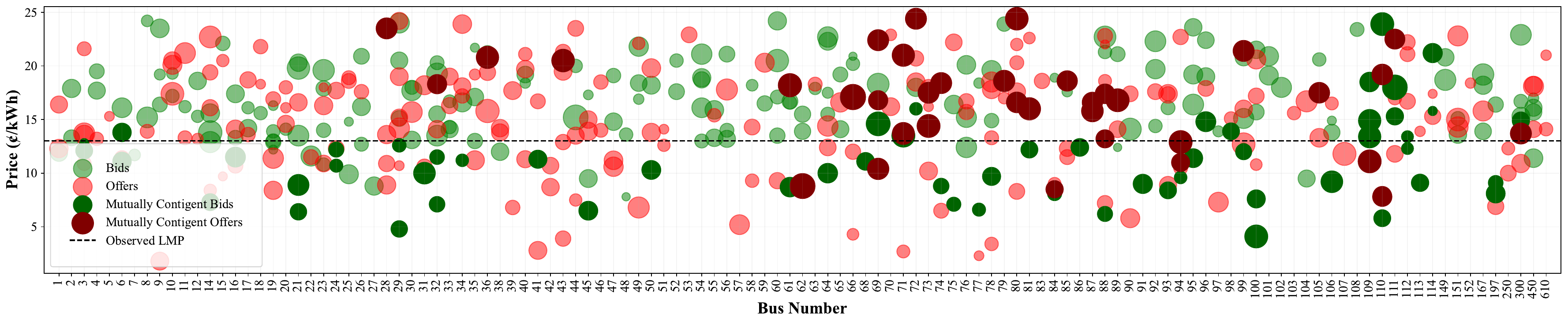}
    \caption{Locations of the randomly generated bids/offers from the Gaussian distributions. The bubble's size reflects the quantum quantity of the bids and offers. The dataset can be viewed at \cite{Dataset}.}
    \label{fig:BidOffers}
\end{figure*}
The head bus (bus $0$) is the linkage point at the \textit{Transmission-Distribution interface}, where the IDSO's substation is located. The base parameter values are set as follows: $S_{\mathrm{base}} = 1000$ kVA, $V_{\mathrm{base}} = 2.401$ kV. The squared node voltage limits are defined by $\bm{v}_{\mathrm{min}} = [0.95^2, 0.95^2, 0.95^2]^T$, $\bm{v}_{\mathrm{max}} = [1.05^2, 1.05^2, 1.05^2]^T$, and the substation voltage is set to $\bm{v}_0 = [1.03^2, 1.03^2, 1.03^2]^T$. The maximum apparent power $S^\phi_{(i,j),\mathrm{max}}$ for each line and phase is derived from its thermal capacity based on ampacity values provided in \cite{Kersting2007}, and the substation apparent power limit is set as $S^\phi_{0,\,\mathrm{max}} = 5000$ kVA across all phases. Fixed loads are modelled at the same locations and with half the magnitudes as in the standard distribution network. The total fixed load in the system is $1347.5$ kW and $960$ kVAr. The bids/offers from the DERs are assumed to follow Gaussian distributions given as \(\mathcal{N}(20,10^2) \in [5,45]\) for bid power (kW), \(\mathcal{N}(-20,10^2) \in [-45,-5]\) for offer power (kW), and \(\mathcal{N}(15,5^2) \in [1,25]\) for bid/offer price (\textcent/kWh). The generated bids/offers, along with their locations, are shown in Fig. \ref{fig:BidOffers}; the generated dataset is available at \cite{Dataset}.
The DERs are assumed to be operated at a constant power factor, $\theta(\psi) = 0.9$. The $M$ used to initialise $\gamma(\psi)$ for offers is taken as $1000$. The IDSO's marginal cost of providing network service is considered as $m= 2.5\text{ ¢/kWh}$. The LMP in the WPM is considered to be set at $13$ ¢/kWh.\nomenclature[A]{\(V_{\mathrm{base}}\)}{Base value for voltage in kV;}\nomenclature[A]{\(\bm{v}_{\mathrm{min}}, \bm{v}_{\mathrm{max}}\)}{3-phase vector for min/max limits on voltage in p.u.;}\nomenclature[A]{\(S^\phi_{(i,j),\mathrm{max}}\)}{Thermal Capacity of line $(i,j)$ and phase $\phi$  in kVA;}\nomenclature[A]{\(S^\phi_{0,\,\mathrm{max}}\)}{IDSO's Substation's apparent power limit for phase $\phi$ in kVA;}
\vspace{-1em}
\subsection{Test Cases and Results}
As previously discussed, TES designs within ITD systems introduce tight T-D linkages that must be accounted for to ensure seamless system operations. Using the generated bids and offers data, shown in Fig. \ref{fig:BidOffers} (available at \cite{Dataset}), the following test cases illustrate the necessity of the proposed methodology to effectively capture these linkages. 
\subsubsection{Aggregation of Bids and Offers without Excluding Mutually Contingent Bids/Offers}
\label{Test:Agg}
This test case highlights the importance of \textit{ex-ante aggregation}, wherein the mutually contingent bids and offers are identified prior to IDSO's participation in WPM. This test case demonstrates how the network may experience constraint violations when these bids and offers are not simultaneously cleared in the WPM. 
\begin{figure}[!htbp]
    \centering
    \subfloat[\label{fig:VoltageContingent}]{
        \includegraphics[width=\linewidth, height=3.1cm]{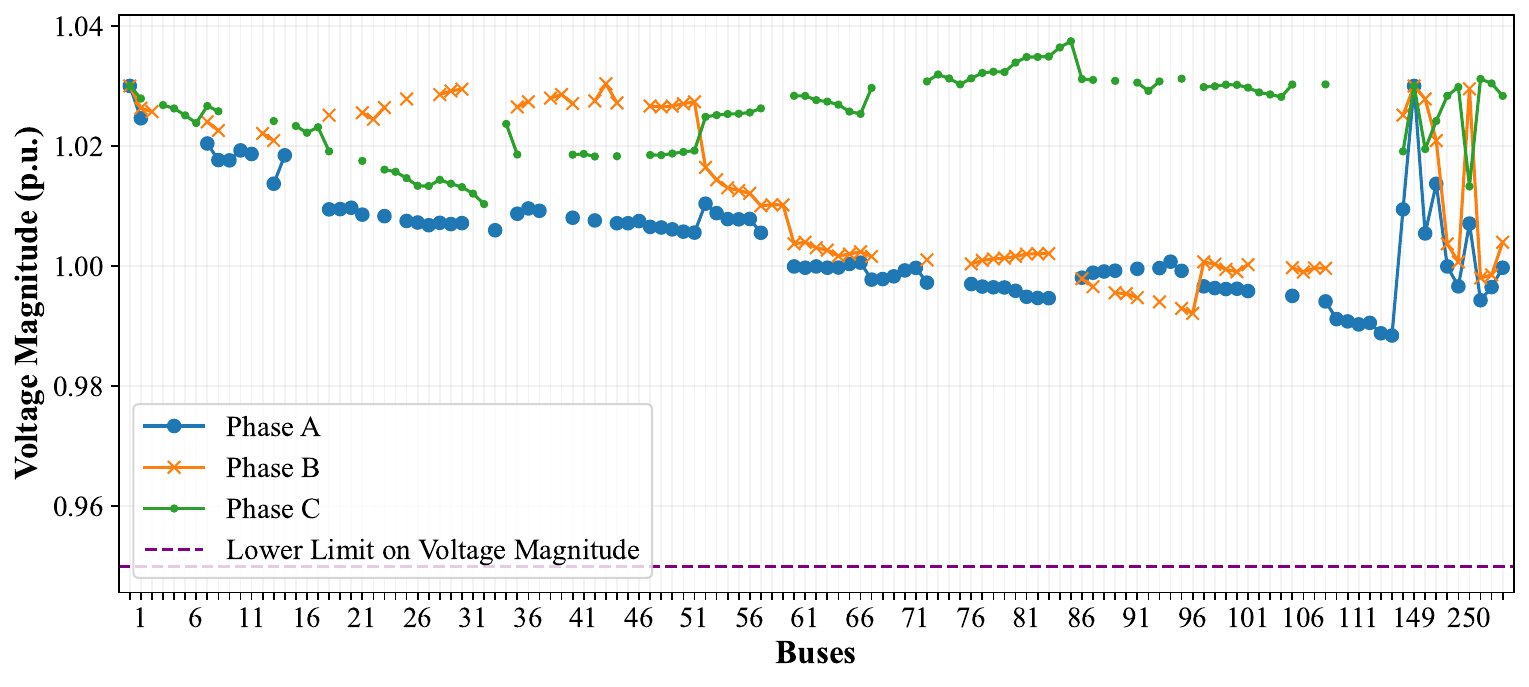}
    }
     
    \subfloat[\label{fig:VoltageNaiveContingent}]{
        \includegraphics[width=\linewidth, height=3.1cm]{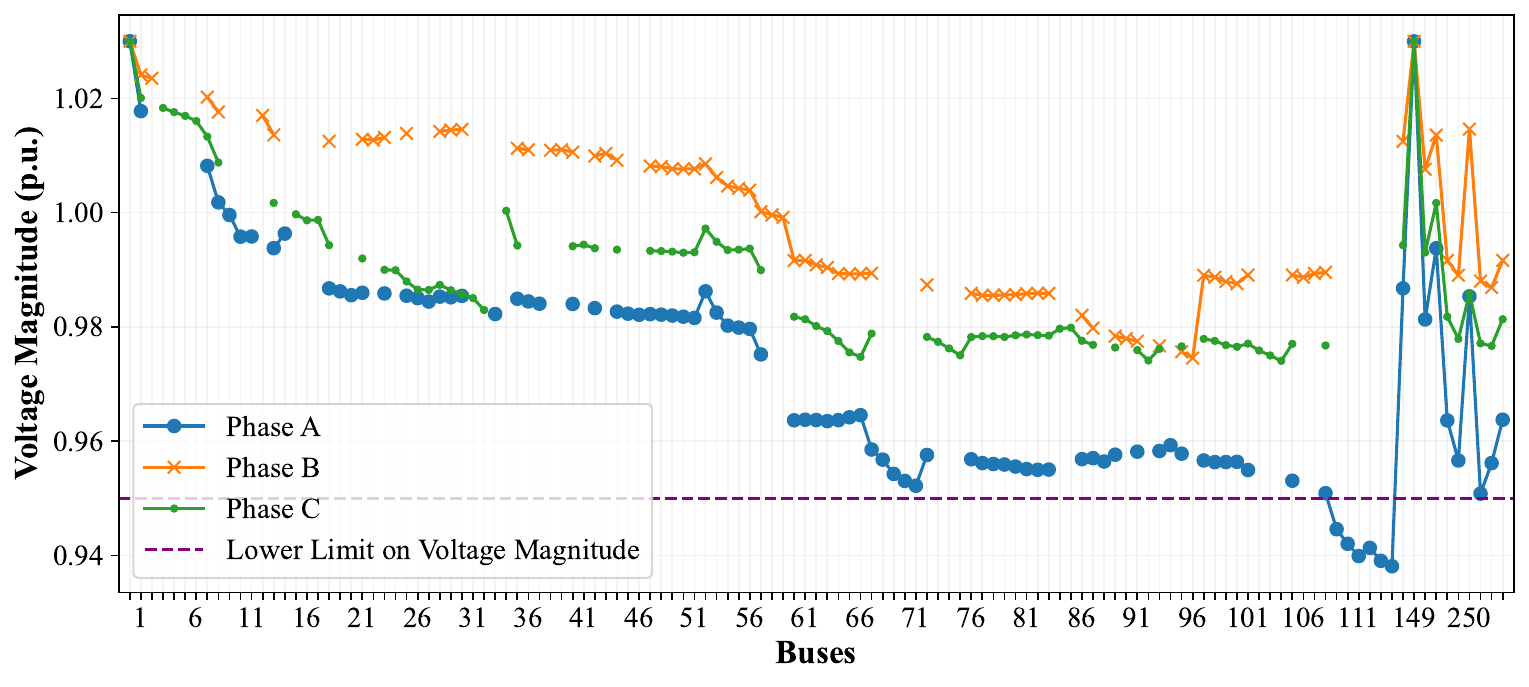}
    }
    \caption{3-phase voltage magnitudes across the distribution network (a) before and (b) after WPM clearing for the Test Case 1.}
    \label{fig:CombinedVoltage}
\end{figure}

The IDSO solves its aggregation problem described in Def.~\ref{BOAProblem} using all bids and offers from the input dataset. The resulting cleared bids and offers are then aggregated for participation in the WPM. Note that because this aggregation ignores mutually contingent bids and offers, all bids and offers from the T-DOPF with $\alpha(\psi) \neq 0$ qualify for WPM participation. When these bids and offers withdraw and inject power, respectively, the resulting three-phase voltage magnitudes across the distribution network are shown in Fig. \ref{fig:VoltageContingent}. As expected from any T-DOPF that considers network constraints, the voltages at all buses remain within acceptable limits. However, given an observed LMP of $13$ \textcent/kWh, only offers with centres below the LMP line in Fig. \ref{fig:BidOffers} and bids with centres above it are cleared in the WPM. Because not all qualified bids and offers are cleared for power injection or withdrawal, the distribution network constraints may get violated, as shown by the resulting three-phase voltage magnitudes plotted in Fig. \ref{fig:VoltageNaiveContingent}, where the phase-a voltage exceeds the permissible limit for buses $109-114$. This occurs because the mutually contingent bids and offers are not cleared simultaneously in the WPM.

To address this, the IDSO must identify these mutually contingent bids and offers using the \textit{ex-ante aggregation} technique discussed in Section~\ref{sec:CaseC} and incorporate their interdependencies into its aggregation process. Their presence is identified by creating Bins A, B, C and $\bm{\Psi}_{MC}$ set from the input dataset. $\bm{\Psi}_{MC}$ is then divided into separate bid and offer sets and superimposed on the input bids and offers dataset, as shown in Fig. \ref{fig:BidOffers}. Because, only offers (at buses \(62, 69, 84, 94, 109, \text{ and } 110\)) and bids (at buses \(6, 69, 71, 96, 98, 109, 110, 111, 112, 114, \text{ and } 300\)) are ultimately cleared from the \(\bm{\Psi}_{MC}\) set, the voltage at buses $109-114$ sags due to their partial clearing from the $\bm{\Psi}_{MC}$ set.
Hence, IDSO needs to implement the proposed methodology in Section~\ref{sec:methodology} to qualify bids/offers for participation in WPM, respecting the interdependence of aggregated bid/offer functions and WPM prices.
\subsubsection{Grid-Safe Aggregation Without Considering Network Costs}
To highlight the importance of accounting for network costs while formulating the IDSO's aggregated bid/offer function, a scenario is presented where IDSO bypasses this step and forwards the aggregated bids and offers directly to WPM without accounting for network costs. It tries to recover its network costs later by setting the retail prices as shown below. 
\par
\begin{figure*}[!htbp]
    \centering
    \subfloat[]{%
        \includegraphics[width=0.35\textwidth, height=3.3cm]{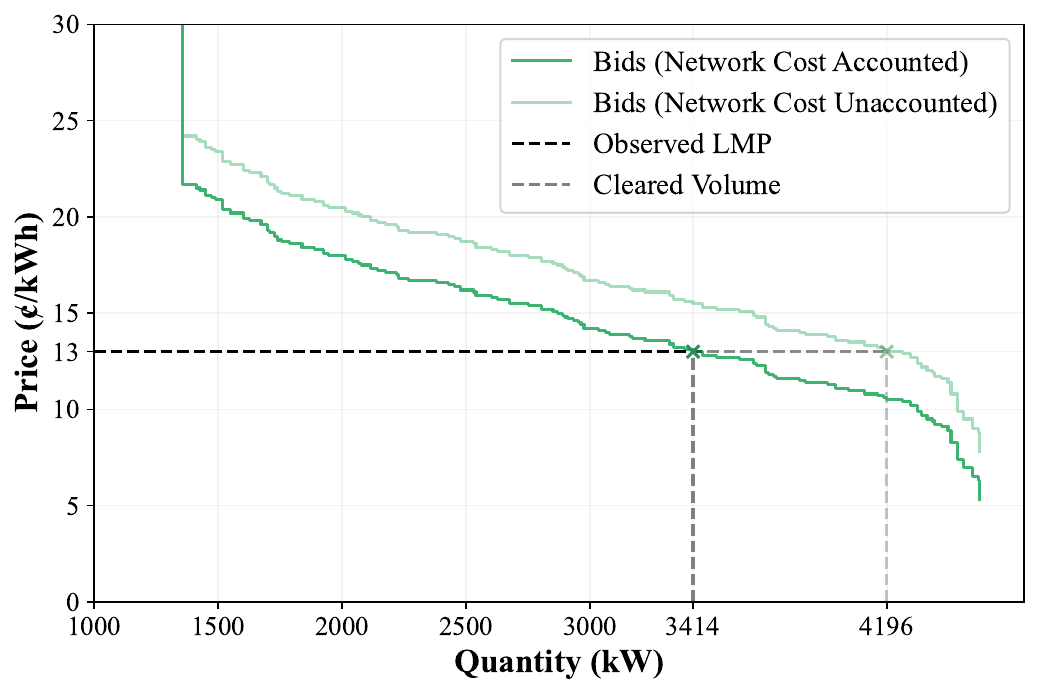}
    }
    \subfloat[]{%
        \includegraphics[width=0.35\textwidth, height=3.3cm]{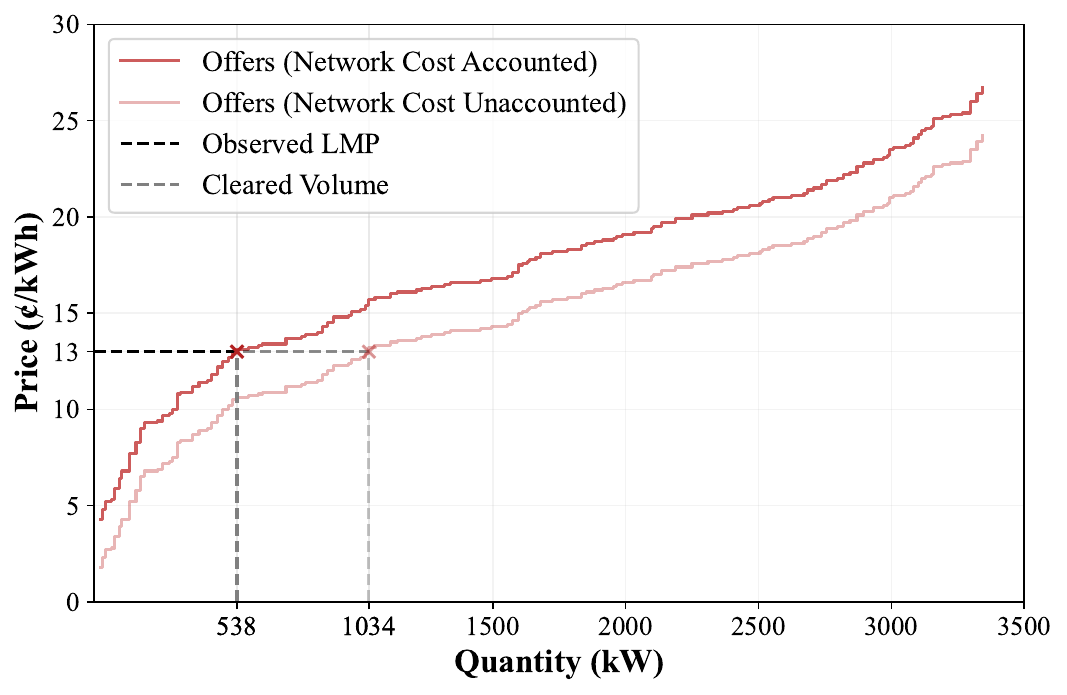}
    }
    \caption{(a) IDSO's bid curve and (b) offer curve, both with and without accounting for network costs, submitted to ITSO for WPM participation.}
    \label{fig:IDSO'sCurves}
\end{figure*}

 For the observed $LMP$ of $13$ ¢/kWh at the IDSO's bus, it sets the retail price as:
\[\pi^{\mathrm{ret}}({\psi}) = 
    \begin{cases}
        13 + 2.5 = 15.5 \text{ ¢/kWh }; &\forall \psi \in \mathcal{C}_{\bm{\Psi}}^{-}, \\
        13 - 2.5 = 10.5 \text{ ¢/kWh }; &\forall \psi \in \mathcal{C}_{\bm{\Psi}}^{+}.
    \end{cases}\]
As bids and offers indicate the reservation values of DERs, a bidding DER cleared in the WPM with a price $\pi(\psi) \in [13,15.5)$ ¢/kWh will not be willing to consume power at a retail price of $\pi^{\mathrm{ret}}({\psi}) = 15.5$ ¢/kWh, as this exceeds its maximum willingness to pay. Similarly, an offering DER cleared in the WPM with a price $\pi(\psi) \in (10.5,13]$ ¢/kWh will not be willing to produce power at a retail price of $\pi^{\mathrm{ret}}({\psi}) = 10.5$ ¢/kWh, since this is below its minimum acceptance price. 
Such situations disrupt tight T-D linkages due to a mismatch between expected and actual net power withdrawal. This can be verified from Fig.~\ref{fig:IDSO'sCurves}, which illustrates the aggregated bids and offers with and without accounting for network costs for the given test system. This underscores the crucial role of the formulation of IDSO's aggregated bid/offer function accounting for the network cost a priori to WPM participation. 
The proposed aggregated bid/offer function ensures that only those DERs capable of covering the associated network costs are qualified for WPM participation, thereby avoiding any contractual power mismatch and allowing tight T-D linkages to be maintained.
\subsubsection{Proposed Framework}
\begin{figure}[!htbp]
    \centering
    \subfloat[\label{fig:BidNQPs}]{%
        \includegraphics[width=\linewidth, height = 4.25cm]{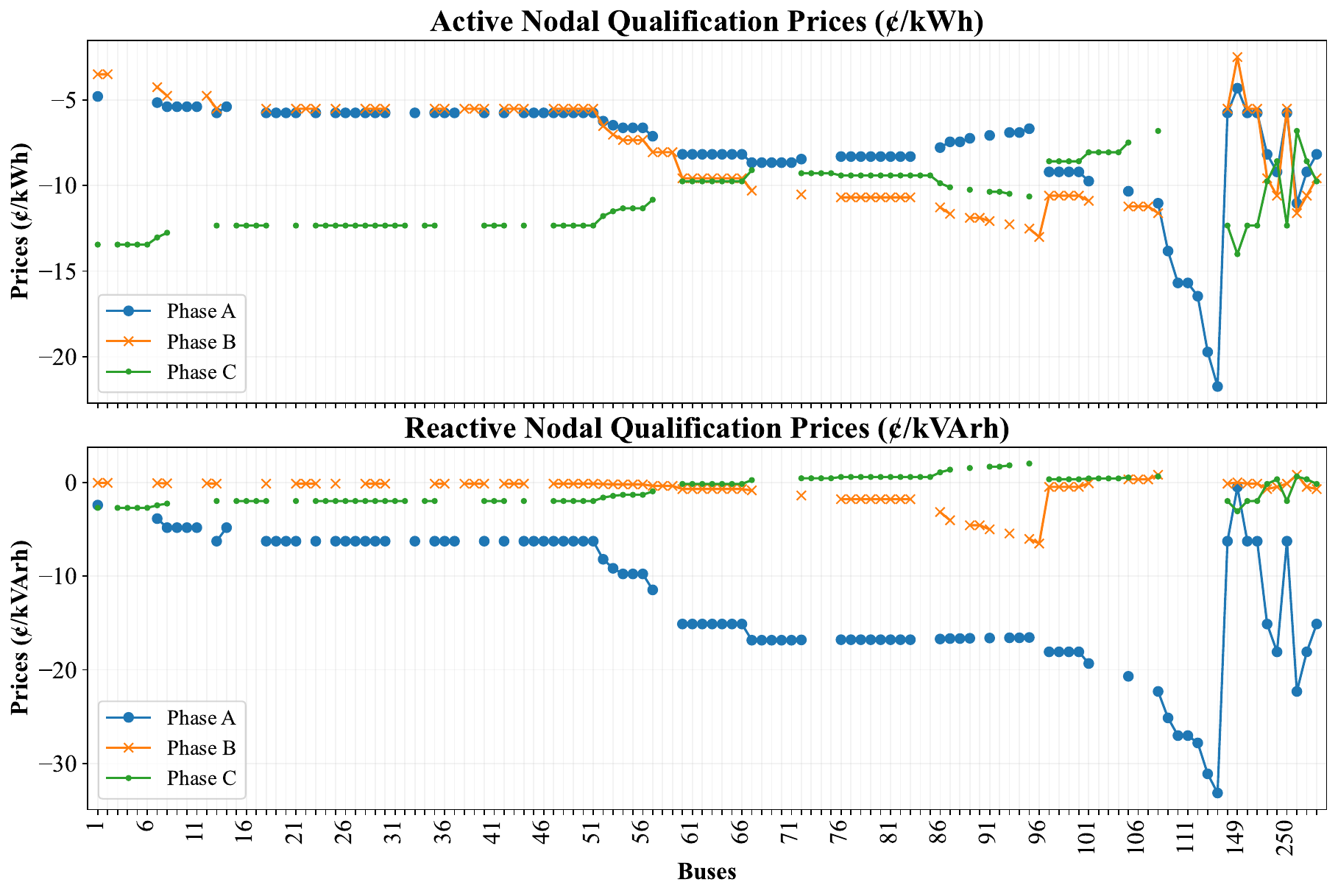}
        }
    \quad
    \subfloat[\label{fig:OfferNQPs}]{%
        \includegraphics[width=\linewidth, height = 4.25cm]{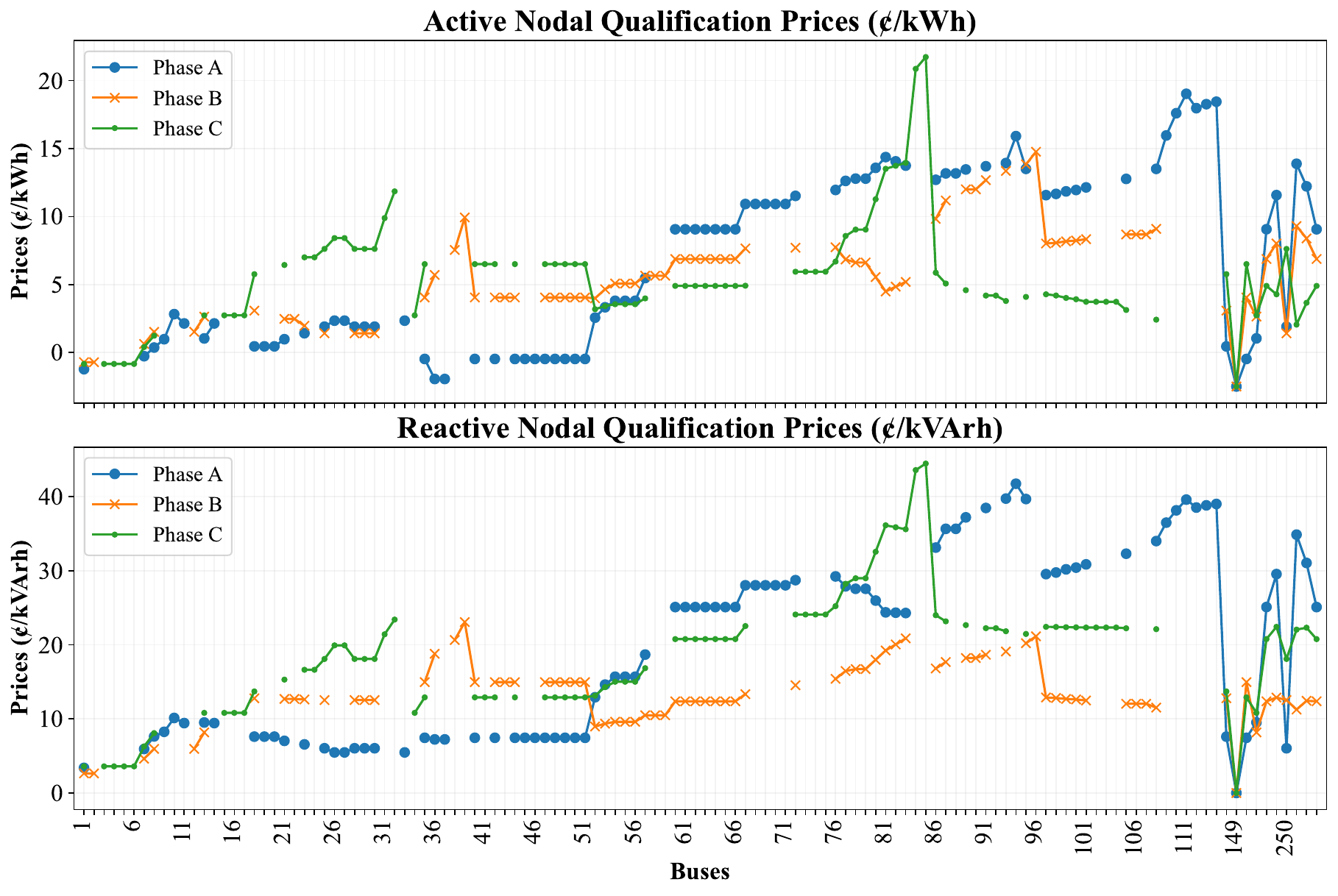}
        }
    \caption{A-NQPs and R-NQPs for the (a) bids in Bin A and (b) offers in Bin B.}
    \label{fig:NQPs}
\end{figure}

\begin{figure}[!htbp]
    \centering
    \includegraphics[width=\linewidth, height=3.7cm]{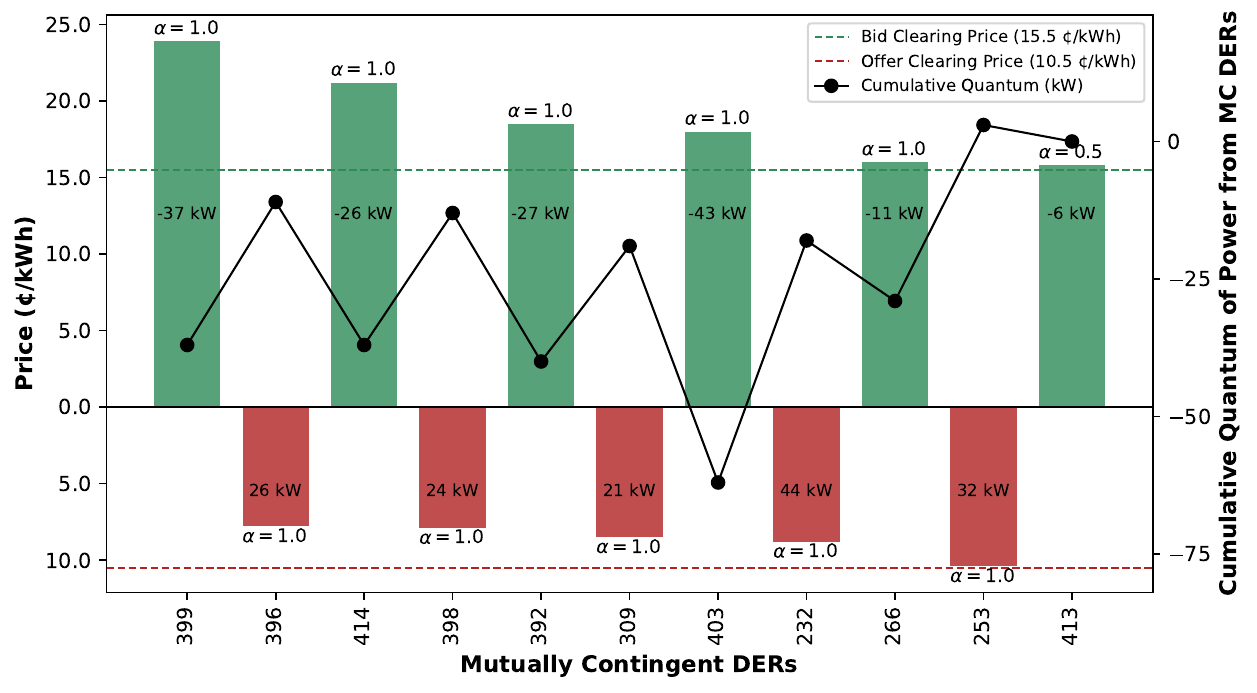}
    \caption{Cleared Mutually Contingent DERs in set $\mathcal{C}_{\bm{\Psi}}^{MC}$.}
    \label{fig:MCBidsOffers}
\end{figure}

\begin{figure}[!htbp]
    \centering
    \includegraphics[width=\linewidth, height=3.4cm]{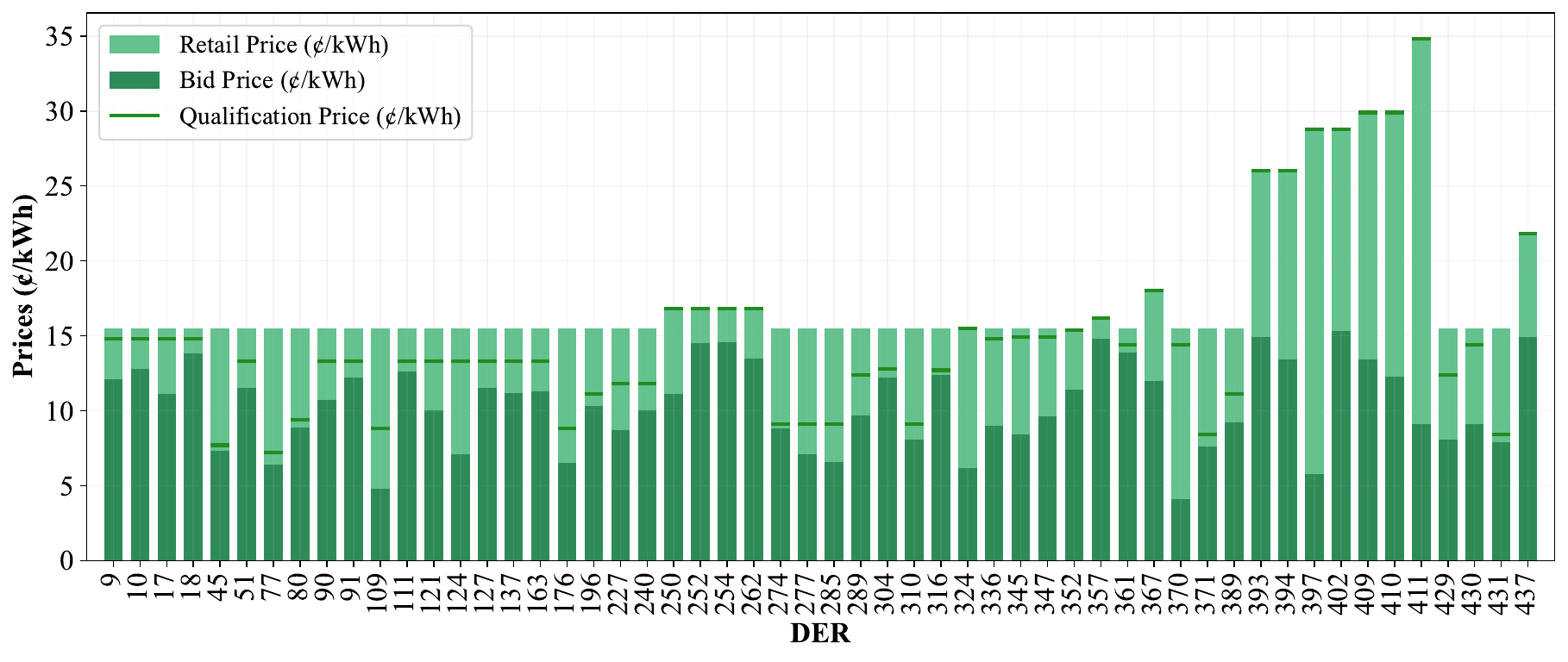}
    \caption{Retail price, bid price and qualification price comparison of unqualified bids.}
    \label{fig:BidsRet}
\end{figure}

\begin{figure}[!htbp]
    \centering
    \includegraphics[width=\linewidth, height=3.4cm]{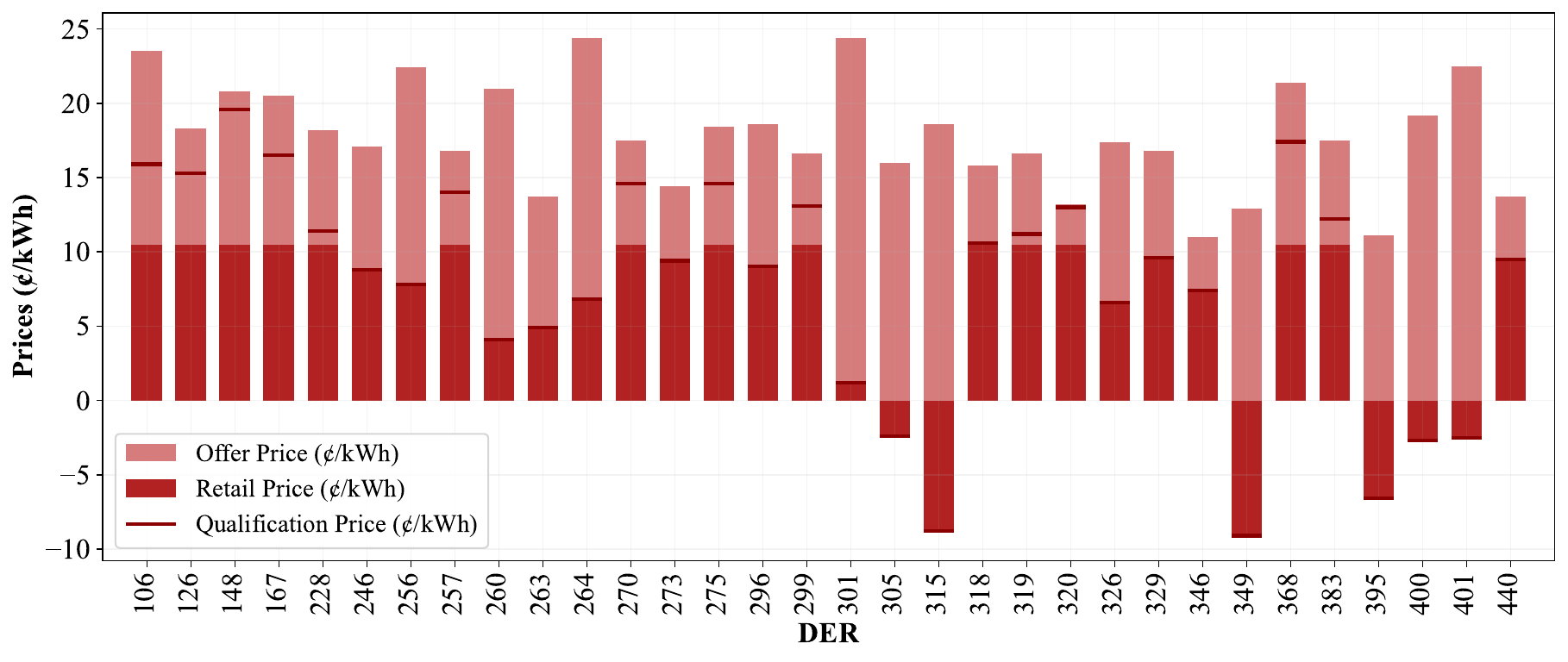}
    \caption{Retail price, offer price and qualification price comparison of unqualified offers.}
    \label{fig:OffersRet}
\end{figure}

\begin{figure}[!htbp]
    \centering
    \includegraphics[width=\linewidth, height=3.1cm]{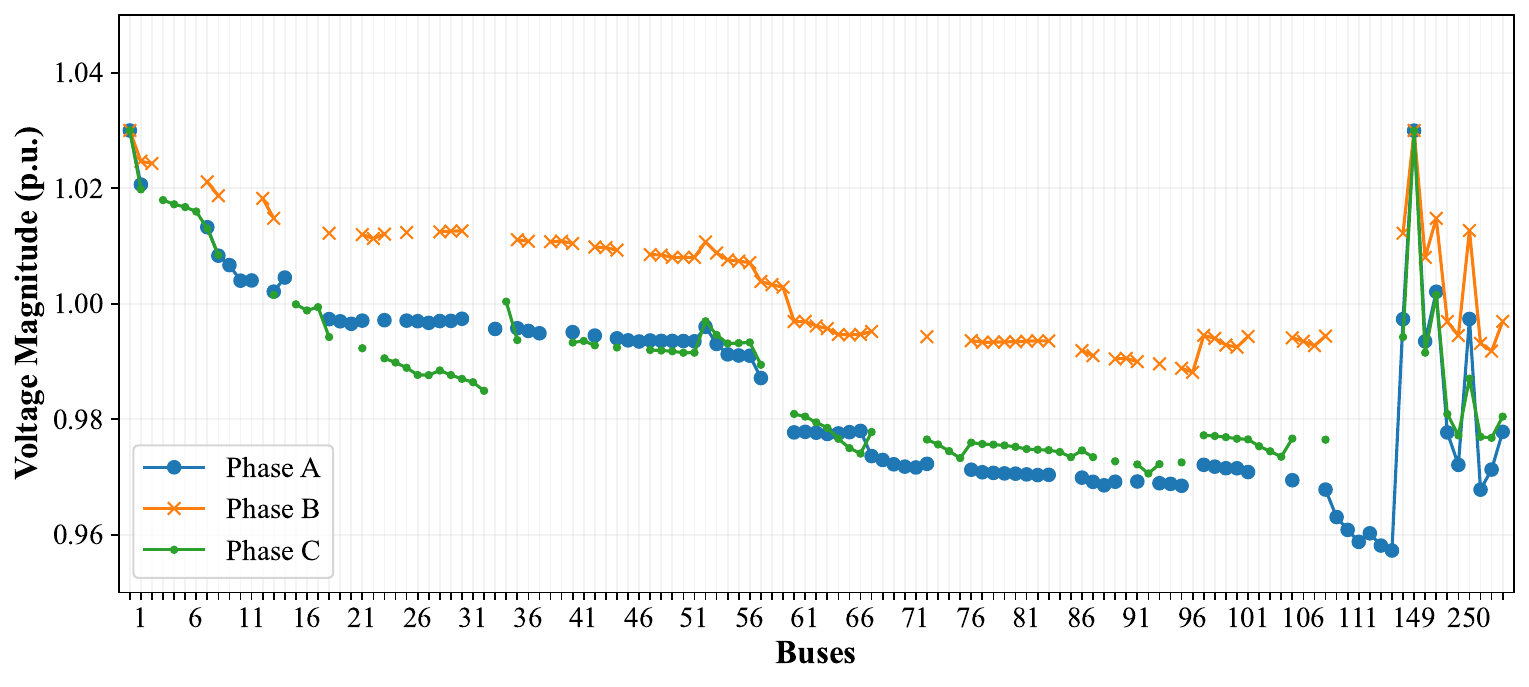}
    \caption{3-phase voltage magnitude across the distribution network following the set retail signals.}
    \label{fig:Final}
\end{figure}

This test case illustrates how the methods described in this study can effectively meet the proposed objectives. As described in Section \ref{sec:CaseC}, IDSO creates Bins A, B and C using T-DOPF on the input dataset. The corresponding A-NQPs and R-NQPs of the T-DOPF for bids in Bin A and offers in Bin B are obtained and shown in Fig. \ref{fig:NQPs}.
Using these NQPs, IDSO applies Prop.~\ref{eitherprop}.\ref{Diffpeitherbid} and Prop.~\ref{eitherprop}.\ref{Diffpeitheroffer} to determine qualified bids and offers, then constructs modified bids and offers using \eqref{eq:IDSO'sBidOfferPrice} and \eqref{eq:IDSO'sBidOfferQuantity}, and aggregates them to participate in the WPM, as illustrated in Fig. \ref{fig:IDSO'sCurves}.
The ITSO clears and determines the LMP of the WPM. As shown in Fig.~\ref{fig:IDSO'sCurves}, the LMP is set at $13$ ¢/kWh, the IDSO's cleared bid volume is $3414$ kW, and the cleared offer volume is $538$ kW. Consequently, the total scheduled net interchange at the T-D linkage bus is $2876$ kW.
\begin{figure}[!htbp]
    \centering
    \subfloat[\label{fig:Voltage(BinI)}]{%
        \includegraphics[width=\linewidth, height=3.1cm]{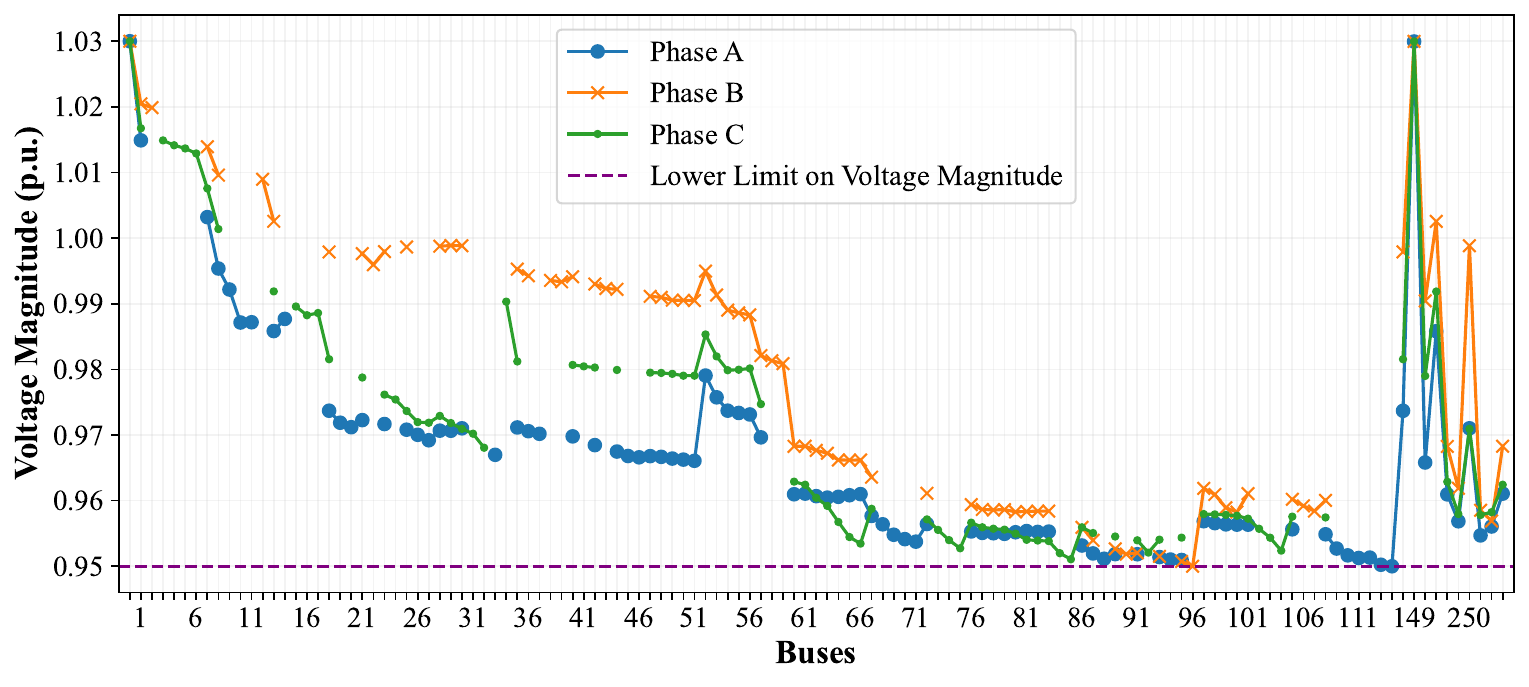}
    }
    \quad
    \subfloat[\label{fig:Voltage(BinII)}]{%
        \includegraphics[width=\linewidth, height=3.1cm]{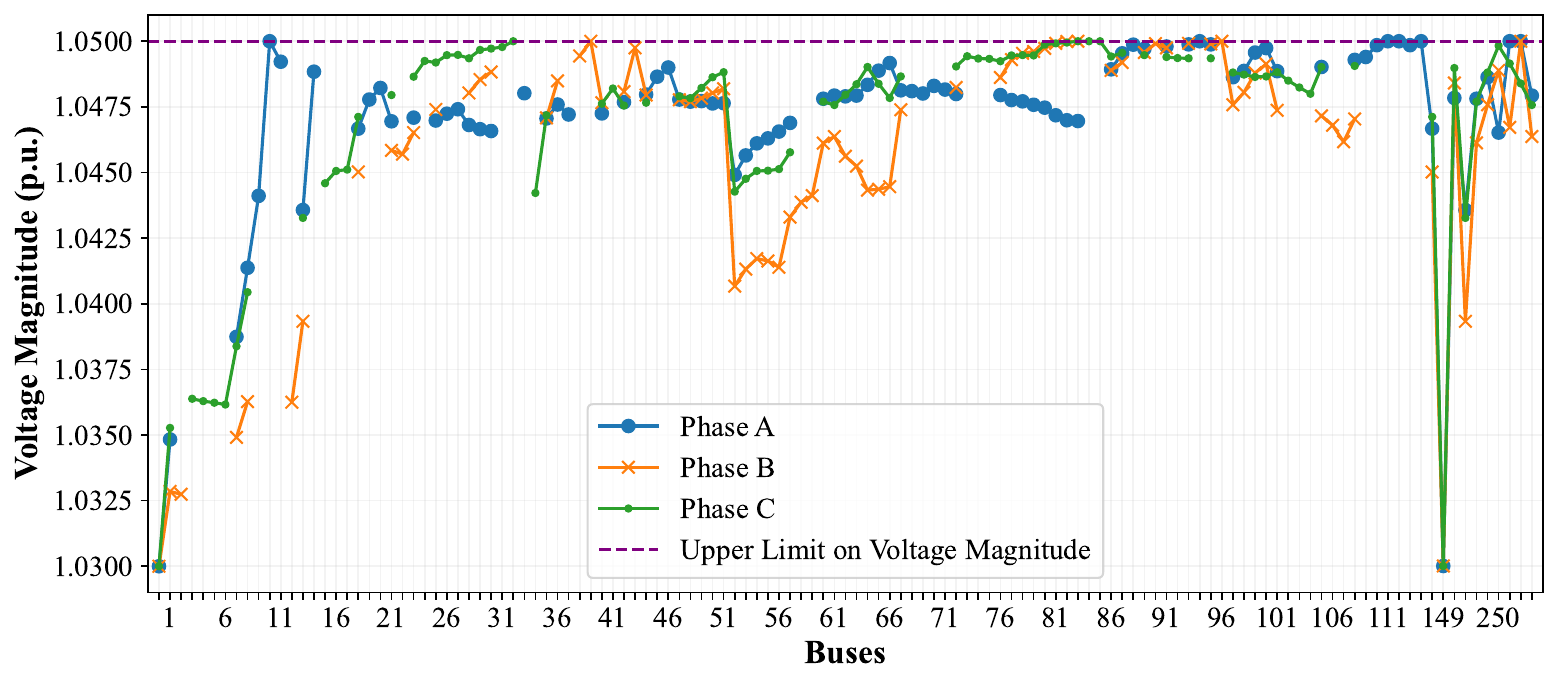}
    }
    \caption{3-phase voltage magnitude across the distribution network for (a) Bin A and (b) Bin B.}
    \label{fig:Voltage(Bins)}
\end{figure}
Once the WPM results are known, the IDSO clears some mutually contingent bids and offers by accounting for network constraints and ensuring that net interchange at the T-D linkage does not exceed the contractual amount. The set of cleared mutually constrained DERs given by \(\mathcal{C}^{MC}_\Psi\) is determined as discussed in Section~\ref{sec:MCBidsOffers}. The corresponding DERs are shown in Fig.~\ref{fig:MCBidsOffers}. 
As is evident, the quantum of cleared bids and offers is equal, and therefore, the IDSO does not violate its contractual agreement with WPM by clearing them. Next, the IDSO must send appropriate retail signals based on the WPM LMP and network costs to the participating DERs, as discussed in Section \ref{sec:RetailDesign}. The IDSO accomplishes this by setting the retail prices using Def.~\ref{def:retail}.
\begin{equation}
    \begin{aligned}
            \pi^\mathrm{ret}(\psi) &=
            \begin{cases}
           \max \, (15.5,\ \ \pi^\mathrm{QP}(\psi)) \text{ \textcent/kWh},&\forall\,\psi \in \overline{\bm{\Psi}}_A,\\
            \min \, (10.5,\ \ \pi^\mathrm{QP}(\psi)) \text{ \textcent/kWh},&\forall\,\psi \in \overline{\bm{\Psi}}_B,\\ 
            15.5 \text{ \textcent/kWh}, &\forall\,\psi \in \bm{\Psi}_A,\\
            10.5 \text{ \textcent/kWh}, &\forall\,\psi \in \bm{\Psi}_B.\\
            \end{cases}
    \end{aligned}
\end{equation}
\begin{equation}
    \begin{aligned}
        p^\mathrm{ret}(\psi) = 
        \begin{cases}
            0, \quad &\forall \psi \in \overline{\bm{\Psi}}_A \cup \overline{\bm{\Psi}}_B,\\
            \alpha(\psi) p(\psi), \quad &\forall \psi \in \bm{\Psi}_A \cup \bm{\Psi}_B.
        \end{cases}
    \end{aligned}
\end{equation}
Figures \ref{fig:BidsRet} and \ref{fig:OffersRet} depict these retail prices for unqualified bids and offers, respectively. Following the set retail signals, the DERs draw/inject power from/to the grid, ensuring the voltage is within operating limits, as depicted in Figure \ref{fig:Final}.
Additional insights on these retail prices and the obtained NQPs in Fig. \ref{fig:NQPs} are discussed below. 
\paragraph{Relationship between NQPs and Constraints}
As outlined in Prop. \ref{prop1bid}, the A-NQPs give the additional markup over IDSO's marginal cost of providing network services that come into effect due to the activation of constraints. The NQPs for bids are negative as they indicate that unit increase in the right-hand side of \eqref{eq:lambda_p} and \eqref{eq:lambda_q} would improve the objective function as $p(\psi) <0$, while for the offers with $p(\psi) > 0$, they are positive, indicating that a unit increase in the right-hand side of \eqref{eq:lambda_p} and \eqref{eq:lambda_q} deteriorates the objective function. For bus $114$, as shown in Fig. \ref{fig:Voltage(BinI)}, the phase-a voltage is at its lower limit of $0.95$ p.u., therefore the corresponding dual variable $\bar{\mu}^v_{114, A}$ will get activated and hence for that node and phase the A-NQP suddenly falls. The same can be said about Bus $96$, where the phase-b voltage hits the lower limit, and the A-NQP for that phase also falls.

Also, as seen in Fig. \ref{fig:BidNQPs}, the R-NQPs for phase-b and phase-c at some buses are positive, indicating that a unit increase in the right-hand side of (\ref{eq:lambda_q}) could result in a less optimal solution. This can be attributed to the fact that more reactive load at phase-b and phase-c at these buses could further deteriorate the voltages, resulting in fewer qualified bids. 
\paragraph{Significance of Negative Retail Prices} 
Some DERs offering power receive negative retail price signals. Such negative signals predominantly occur at leaf nodes (e.g., buses $85, 94,$ and $111$) and far-end buses ($109$ and $110$). This is due to voltage constraint activation at these buses, as shown in Fig. \ref{fig:Voltage(BinII)}, where the voltage is at the upper limit, which makes the corresponding dual variable, $\bar{\mu}_v$, to become active. 
Consider DER 315, connected across phase-c with \(p({315}) = 29\) kW and \(\pi({315}) = 18.6\) \textcent/kWh \cite{Dataset}. From Fig.~\ref{fig:OfferNQPs}, the dual variables are \(\lambda^p_{85,c} = 21.72\) \textcent/kWh and \(\lambda^q_{85,c} = 44.46\) \textcent/kVArh. Assuming a power factor of 0.9, the qualification price is
\[
\pi^\mathrm{QP}({315}) = [-21.72 - (0.484 \times 44.46)] + \frac{1000}{29} \approx -8.8 \text{ \textcent/kWh}.
\]
Consequently, \(\pi^\mathrm{ret}({315})\) is \(-8.8\) \textcent/kWh, matching the value shown in Fig.~\ref{fig:OffersRet}. 
This result, similar to the negative LMPs seen in WPM, acts as a signal for the IDSO to consider installing voltage support devices to allow the grid to accommodate further power injections.
\section{Conclusion}
\label{sec:conclusion}
This study presents an IDSO-managed bid-based TES design of an unbalanced distribution system operating within an ITD paradigm. The IDSO aggregates the bids and offers from participating DERs and uses the aggregated bid and offer function to participate in the WPM, while ensuring network reliability of the distribution system. Since the participation of DERs in bid-based designs enables tightly coupled T-D interactions, the proposed framework presents two complementary components designed to preserve these dependencies and capture the dynamics of a realistic ITD system. The first component enables grid-safe and value-based aggregation of DERs using a novel bid/offer prequalification-cum-aggregation method based on T-DOPF to build aggregated bid/offer functions for the IDSO, facilitating WPM participation while preserving the dependencies in the exchanges between IDSO and ITSO. The second component is a retail pricing mechanism designed to facilitate value-based exchanges between the IDSO and DERs. It captures the true value of procuring or offering additional units of power within the distribution system, while also enabling DERs with adequate information to effectively update their bids and offers for subsequent time periods. In conclusion, the design is developed such that the inherent dependencies of the interactions by the IDSO with both the DERs and the ITSO are preserved. Case studies conducted on a modified IEEE 123-bus radial feeder populated with a high DER concentration validate the proposed framework's effectiveness in coordinating the DERs efficiently and reliably.
\appendices
\printnomenclature 
\section{Formulation of Three-Phase Distribution Power Flow}
\label{Append:DistFlowModel}
This study considers an $N+1$ bus unbalanced radial distribution network, characterized by a graph $\mathcal{G}(\mathcal{N} \cup \{0\},\mathcal{L})$, with phases $\phi \in \bm{\Phi}: \{a,b,c\}$. The index of the head bus in the feeder where the IDSO's substation is located is denoted by $0$, and all other non-head buses are the elements of the set $\mathcal{N}: \{1,2, \cdots, N\}$. The distribution network has $N$ distinct line segments denoted by set $\mathcal{L} \subseteq \{\mathcal{N} \cup \{0\}\} \times \{\mathcal{N} \cup \{0\}\}$, with $(i,j) \in \mathcal{L}$ denoting a line segment between buses $i$ and $j$. For each bus $i$ with phase $\phi$, the squared voltage magnitude is denoted by $v_i^\phi$ and the real and reactive power injection by $p_i^\phi$ and $q_i^\phi$, respectively. For a line $(i,j)$ with phase $\phi$, the real and reactive power flow are denoted by $P^\phi_{(i,j)}$ and $Q^\phi_{(i,j)}$, respectively. The resistance and reactance of the line $(i,j)$ between phase $\phi$ of node $i$ and $\phi^`$ of node $j$ is given by $R^{\phi\phi^`}_{(i,j)}$ and $X^{\phi\phi^`}_{(i,j)}$ respectively.
The node voltages, injections and line flows can easily be represented as $3 \times 1$ column vectors as shown in (\ref{eqn:vectors}), where all items are in p.u.\nomenclature[V]{\(v^\phi_i\)}{Squared voltage magnitude of a bus $i$ and phase $\phi$ in p.u.;}\nomenclature[A]{\(R^{\phi,\phi^`}_{(i,j)}, X^{\phi,\phi^`}_{(i,j)}\)}{Resistance and Reactance between phase $\phi$ of bus $i$ and phase $\phi^`$ of node $j$ in p.u.;}\nomenclature[A]{\(P^\phi_{(i,j)}, Q^\phi_{(i,j)}\)}{Real and Reactive power flow in a line \((i,j)\) with phase $\phi$ in p.u.;} 
\begin{subequations}
\allowdisplaybreaks
    \label{eqn:vectors}
    \begin{align*}
        \bm{v}_i = [v_i^{\phi}]_{\phi \in \bm{\Phi}};\quad
        \bm{p}_i =& [p_i^{\phi}]_{\phi \in \bm{\Phi}};\quad
        \bm{q}_i = [q_i^{\phi}]_{\phi \in \bm{\Phi}};\\
        \bm{P}_{(i,j)} = [P_{(i,j)}^{\phi}]_{\phi \in \bm{\Phi}};&\quad
        \bm{Q}_{(i,j)} = [Q_{(i,j)}^{\phi}]_{\phi \in \bm{\Phi}}.
    \end{align*}\nomenclature[A]{\(p_i^\phi, q_i^\phi\)}{Real and reactive power injection at a bus $i$ and phase $\phi$ in p.u.;}\nomenclature[V]{\(\bm{v}_i\)}{3 phase squared voltage column vector for a bus \(i\) in p.u.;}\nomenclature[V]{\(\bm{p}_i, \bm{q}_i\)}{3-phase real and reactive power injection (p.u.) column vector of a bus $i$;}\nomenclature[V]{\(\bm{P}_{(i,j)}, \bm{Q}_{(i,j)}\)}{3-phase real and reactive power flow (p.u.) column vector of a line \((i,j)\);}
\end{subequations}
Also, line resistances and reactances in p.u. for a line ($i,j$) can be equivalently represented in a $3 \times 3$ matrix form as: 
\begin{align*}
        \bm{R}_{(i,j)} = [R_{(i,j)}^{\phi\phi^`}]_{\phi, \phi^` \in \bm{\Phi}}; \quad
        \bm{X}_{(i,j)} = [X_{(i,j)}^{\phi\phi^`}]_{\phi, \phi^` \in \bm{\Phi}} . 
    \end{align*}\nomenclature[A]{\(\bm{R}_{(i,j)}, \bm{X}_{(i,j)}\)}{3-phase resistance and reactance matrix of a line \((i,j)\) in p.u.;}
Then the LinDistFlow equation from \cite{BaranWu} can be represented as follows, where node $k$ is the parent of node $i$,  $\forall k, i \in \mathcal{N},  \forall (i \rightarrow j) \in \mathcal{L}$:
\begin{subequations}
 \label{eqn:LinDistFlow}
 \begin{equation}
     \begin{aligned}
          \sum_{j : i \rightarrow j} \bm{P}_{(i,j)} &= \bm{P}_{(k,i)} + \bm{p}_i        \,\,,
     \end{aligned}
 \end{equation}
 \begin{equation}
     \begin{aligned}
          \sum_{j : i \rightarrow j} \bm{Q}_{(i,j)} &= \bm{Q}_{(k,i)} + \bm{q}_i \,\,,
     \end{aligned}
 \end{equation}
 \begin{equation}
     \begin{aligned}
         \bm{v}_i - \bm{v}_j &= 2 \cdot (\Bar{\bm{R}}_{(i,j)} \bm{P}_{(i,j)} + \Bar{\bm{X}}_{(i,j)}\bm{Q}_{(i,j)}).
     \end{aligned}
 \end{equation}
 \end{subequations}
Where, 
\begin{align*}
        \bar{\bm{R}}_{(i,j)} &= \Re(\bm{W}) \odot \bm{R}_{(i,j)} + \Im(\bm{W}) \odot \bm{X}_{(i,j)}, \\[5pt]
        \bar{\bm{X}}_{(i,j)} &= \Re(\bm{W}) \odot \bm{X}_{(i,j)} - \Im(\bm{W}) \odot \bm{R}_{(i,j)}.
\end{align*}\nomenclature[A]{$\bar{\bm{R}}_{(i,j)},\bar{\bm{X}}_{(i,j)}$}{3-phase phase coupled resistance and reactance matrix of a line $(i,j)$ in p.u.;}

Here, $\bm{W}$ is the phase coupling matrix:
\begin{equation}
    \begin{aligned}
      &\quad \quad \quad \quad \quad\bm{W} = \begin{bmatrix}
        1 & \omega & \omega^2 \\[5pt]
        \omega^2 & 1 & \omega \\[5pt]
        \omega & \omega^2 & 1
    \end{bmatrix}; \quad \omega = e^{j\frac{2\pi}{3}},\\
        &\odot \text{ denotes the Hadamard Product}. \notag  
    \end{aligned}
\end{equation}\nomenclature[A]{\(\bm{W}\)}{Phase coupling matrix;}\nomenclature[A]{\(\omega=e^{j\frac{2\pi}{3}}\)}{Phase shift operator;}
Using \cite{lowlecture_notes}, the above linear power flow equations can be conveniently represented by a graph-based matrix formalism. Let the branch-by-node incidence matrix for $\mathcal{G}$ be denoted by $\bm{\Tilde{C}} \in \{-\bm{I}^3, 0, \bm{I}^3\}^{(N) \times (N+1)}$ representing a three-phase connection structure of the lines and buses, the entries can then be expressed as $\forall i \in \{\mathcal{N} \cup \{0\}\}, \forall l \in \mathcal{L}$:
\begin{equation}
    \bm{\Tilde{C}}_{il} = 
    \begin{cases} 
    \bm{I}^3, & \text{if line } l \text{ originates from node } i,\\
    -\bm{I}^3, & \text{if line } l \text{ feeds node }  i,\\
    0, & \text{otherwise}. 
    \end{cases}
\end{equation}\nomenclature[A]{\(\bm{\Tilde{C}}\)}{Branch-node incidence matrix for a 3-phase radial distribution network;}
Where $\bm{I}^3$ is an identity matrix of size 3. The LinDistFlow equations in (\ref{eqn:LinDistFlow}) can therefore be represented as:
\begin{subequations}
\allowdisplaybreaks
\label{eqn:MatrixLinDistFlow}
    \begin{align}
    \Tilde{\bm{C}}^T\bm{P} &= \bm{p}; \quad \Tilde{\bm{C}}^T\bm{Q} = \bm{q};\\
        \Tilde{\bm{C}} \bm{v}
        &= 2 \left[ \bm{D_r} \bm{P} + \bm{D_x} \bm{Q} \right].
    \end{align}
\end{subequations}
where, \(\bm{v}, \bm{p}, \bm{q} \in \mathbb{R}^{3(N+1) \times 1}, \text{ and } \bm{P}, \bm{Q} \in \mathbb{R}^{3N \times 1}\) defined as \[
\begin{aligned}
    \bm{v} = \big[\bm{v}_i\big]_{i \in \mathcal{N} \cup \{0\}},\quad
    \bm{p} = \big[\bm{p}_i\big]_{i \in \mathcal{N} \cup \{0\}}, \quad 
    \bm{q} = \big[\bm{q}_i\big]_{i \in \mathcal{N} \cup \{0\}}, \\
    \bm{P} = \big[\bm{P}_{(i,j)}\big]_{(i,j) \in \mathcal{L}}, \quad 
    \bm{Q} = \big[\bm{Q}_{(i,j)}\big]_{(i,j) \in \mathcal{L}}.
\end{aligned}
\]\nomenclature[V]{\(\bm{v}\)}{3-phase square voltage magnitudes (p.u.) column vector of all buses in the distribution network;}\nomenclature[V]{\(\bm{p,q}\)}{3-phase real and reactive power injections (p.u.) column vector across all buses in the distribution network;}
$\bm{D_r}$ and $\bm{D_x} \in \mathbb{R}^{3N \times 3N}$  are diagonal matrices, defined as:
\begin{align*}
        \bm{D_r} &= \text{ diag}((\Bar{\bm{R}}_{(i,j)})_{(i,j) \in \mathcal{L}}), \\
        \bm{D_x} &= \text{ diag}((\Bar{\bm{X}}_{(i,j)})_{(i,j) \in \mathcal{L}}).
\end{align*}
Let the first column of $\bm{\Tilde{C}}$ be denoted by the column vector $\bm{c_0}$ that corresponds to the head bus, while the rest of the matrix as $\bm{C}$. Therefore, the incidence matrix can be represented more compactly as:
\begin{equation}
    \label{eqn:incidence}
    \bm{\Tilde{C}} = \begin{bmatrix}
                        \bm{c}_0 & \bm{C}
                     \end{bmatrix}.
\end{equation}
Therefore, the equation (\ref{eqn:MatrixLinDistFlow}) can be written as:
\begin{subequations}
\allowdisplaybreaks
\label{eqn:MatrixLinDistFlowFinal}
    \begin{align}
    \label{eqn:eqn:MatrixLinDistFlowFinalp0q0}
    \bm{c}^T_0\bm{P} &= \bm{p}_0; \quad \bm{c}^T_0\bm{Q} = \bm{q}_0;\\
    \label{eqn:eqn:MatrixLinDistFlowFinalpq}
    \bm{C}^T\bm{P} &= \bm{p}_{1:N}; \quad \bm{C}^T\bm{Q} = \bm{q}_{1:N};\\
    \label{eqn:eqn:MatrixLinDistFlowFinalV}
    \bm{c}_0\bm{v}_0 + \bm{C} \bm{v}_{1:N}
        &= 2 \left[ \bm{D_r} \bm{P} + \bm{D_x} \bm{Q} \right].
    \end{align}
\end{subequations}
Since, $\bm{\Tilde{C}}$ is invertible \cite{Zhu_OPF},(\ref{eqn:eqn:MatrixLinDistFlowFinalV}) can be written as
\begin{align}
    \label{eqn:MatrixLinDistFlowVFinal}
         \bm{v}_{1:N}
        &= -\bm{C}^{-1} \bm{c}_0\bm{v}_0 + 2 \bm{C}^{-1} \left[ \bm{D_r} \bm{P} + \bm{D_x} \bm{Q} \right]. 
\end{align}
For a tree topology network, the following holds\footnote{Here, $\bm{I}^3_{N+1}$ denotes a column vector of size $N+1$ with individual entries being $\bm{I}^3$},
\begin{subequations}
    \begin{align}
      \Tilde{\bm{C}} \bm{I}^3_{N+1} &= \bm{0}, \\ 
      \bm{c}_0 \bm{I}^3+ \bm{C}\bm{I}^3_N &= 0,\\
      \label{eqn:Cinverse}
      \bm{I}^3_N &= -\bm{C}^{-1}\bm{c}_0.
    \end{align}
\end{subequations}
Using (\ref{eqn:eqn:MatrixLinDistFlowFinalpq}) and (\ref{eqn:Cinverse}), (\ref{eqn:MatrixLinDistFlowVFinal}) can be re-written as:
\begin{align}
    \label{eqn:MatrixLinDistFlowVFinal2}
        \bm{v}_{1:N}
        &= \bm{I}^3_N\bm{v}_0 + 2 \bm{C}^{-1} \left[ \bm{D_r} \bm{P} + \bm{D_x} \bm{Q} \right]. 
\end{align}
\section{Proof of Proposition \ref{prop1bid}}
\label{Append:Prop1}
\begin{IEEEproof}
The lagrangian of the function in (\ref{eq:FullOPF}) is given by:   
\begin{equation}
    \label{eq:lagrangian}
    \begin{aligned}
        &L(\bm{P, Q, \alpha_\Psi, \lambda, \mu}) = \bm{\gamma}^T_\Psi\bm{\alpha}_\Psi\bm{p}_\Psi S_\mathrm{base}\Delta t + C_\mathrm{IDSO}(\bm{P}) \\&+ \bm{\lambda}^{{p}} [\bm{p}^f+\bm{p}^\Psi-\bm{C}^T\bm{P}] + \bm{\lambda}^{{q}} [\bm{q}^f+\bm{q}^\Psi-\bm{C}^T\bm{Q}] \\&+ \underline{\bm{\mu}}^{{v}}[\bm{v}_{min} - \bm{I}^3_N\bm{v}_0 - \bm{v}_P] + \overline{\bm{\mu}}^{{v}}[\bm{I}^3_N\bm{v}_0 + \bm{v}_P - \bm{v}_{max}] \\&+
        \sum_{e\in \mathcal{E}^P}\bigl[\bm{\mu}^{{P}}(e)[\beta_e\bm{P} + \delta_e\bm{Q}+\gamma_eS_{max}]\bigr] \\&+
        \sum_{e\in \mathcal{E}^{sub}}\bigl[\bm{\mu}^{{sub}}(e)[\beta_e\bm{p}_0 + \delta_e\bm{q}_0+\gamma_eS_{0,max}]\bigr].
    \end{aligned}
\end{equation}
The Karush-Kuhn-Tucker Conditions, at optimality yield $ \nabla_{\bm{\alpha}_\Psi} L^*=0,\nabla_{\bm{P}}L^*=0, \nabla_{\bm{Q}}L^*=0$, from these we obtain---
\begin{equation}
\label{eq:28}
    \begin{aligned}
        \bm{p}_\Psi^T\bm{\gamma}_\Psi S_\mathrm{base}\Delta t = -\bm{p}_\Psi^T \bm{A}_\Psi^T\bm{\lambda}^{{p}^{*T}}- \bm{p}_\Psi^T \bm{\eta}_\Psi^T \bm{A}_\Psi^T\bm{\lambda}^{{q}^{*T}}.
    \end{aligned}
\end{equation}
Since $\bm{A}_\psi \text{ and }\bm{\eta}_\psi$ are both diagonal matrices, \eqref{eq:28} becomes:
\begin{equation}
    \label{eq:Prop2final}
    \begin{aligned}
        \bm{p}_\Psi^T\bm{\gamma}_\Psi S_\mathrm{base}\Delta t = \bm{p}_\Psi^T \bm{A}_\Psi^T[-\bm{\lambda}^{{p}^{*T}} - \bm{\eta}_\Psi^T\bm{\lambda}^{{q}^{*T}}].
    \end{aligned}
\end{equation}
Assume a DER $\psi$ is connected bus $i$, (\ref{eq:Prop2final}) becomes:
    \begin{equation}
        \begin{aligned}
            &\begin{bmatrix}
                p^a(\psi)& p^b(\psi)& p^c(\psi)
            \end{bmatrix}
            \begin{bmatrix}
                \gamma(\psi)\\
                \gamma(\psi)\\
                \gamma(\psi)
            \end{bmatrix} S_\mathrm{base}\Delta t \\&= 
            \begin{bmatrix}
                p^a(\psi)& p^b(\psi)& p^c(\psi)
            \end{bmatrix}
            \begin{bmatrix}
               1& 0& 0\\
               0&1&0\\
               0&0&1
            \end{bmatrix}\Biggl[
            \begin{bmatrix}
                -\lambda_{i,a}^p\\
                -\lambda_{i,b}^p\\
                -\lambda_{i,c}^p
            \end{bmatrix} \\&+
            \begin{bmatrix}
               \eta(\psi)& 0& 0\\
               0&\eta(\psi)&0\\
               0&0&\eta(\psi)
            \end{bmatrix}
            \begin{bmatrix}
                -\lambda_{i,a}^q\\
                -\lambda_{i,b}^q\\
                -\lambda_{i,c}^q
            \end{bmatrix}
        \Biggr].
        \end{aligned}
    \end{equation}
Linearly, this can be expressed as:
\begin{equation}
\label{eq:LinearA}
    \begin{aligned}
        &\gamma(\psi)[p^a(\psi)+p^b(\psi)+p^c(\psi)]S_\mathrm{base}\Delta t=\\&-p^a(\psi)[\lambda_{i,a}^p+\eta(\psi)\lambda_{i,a}^q]-p^b(\psi)[\lambda_{i,b}^p+\eta(\psi)\lambda_{i,b}^q]\\&-p^c(\psi)[\lambda_{i,c}^p+\eta(\psi)\lambda_{i,c}^q].
    \end{aligned}
\end{equation}
Using (\ref{eq:PerPhase}), (\ref{eq:LinearA}) can be rewritten as:
\begin{equation}
\label{eq:gamma_psi}
    \begin{aligned}
        \gamma(\psi) = \frac{\sum_{\phi \in \bm{\Phi}(\psi)}-\lambda_{i,\phi}^p + \eta(\psi) \sum_{\phi \in \bm{\Phi}_\psi} -\lambda_{i,\phi}^q}{N_{\phi_\psi} S_\mathrm{base}\Delta t}.
    \end{aligned}
\end{equation}
As $\psi$ is a bidding DER, $\gamma(\psi) = \pi(\psi)$, the previous equation becomes
\begin{equation}
    \begin{aligned}
        \pi(\psi) = \frac{\sum_{\phi \in \bm{\Phi}(\psi)}-\lambda_{i,\phi}^p + \eta(\psi) \sum_{\phi \in \bm{\Phi}_\psi} -\lambda_{i,\phi}^q}{N_{\phi_\psi} S_\mathrm{base}\Delta t};\notag
    \end{aligned}
\end{equation}
Economic principles dictate that a bidding DER would consume only when its bid price is greater than or equal to the marginal value of procuring power at the connected location.
Therefore, the following equation holds for bids:
\begin{equation}
    \begin{aligned}
        \pi(\psi) \geq \frac{\sum_{\phi \in \bm{\Phi}(\psi)}-\lambda_{i,\phi}^p + \eta(\psi) \sum_{\phi \in \bm{\Phi}_\psi} -\lambda_{i,\phi}^q}{N_{\phi_\psi} S_\mathrm{base}\Delta t}.
    \end{aligned}
\end{equation}
Hence, the qualification price is set as
\begin{align}
    \pi^\mathrm{QP}(\psi) \coloneq \frac{\sum_{\phi \in \bm{\Phi}(\psi)}-\lambda_{i,\phi}^p + \eta(\psi) \sum_{\phi \in \bm{\Phi}_\psi} -\lambda_{i,\phi}^q}{N_{\phi_\psi} S_\mathrm{base}\Delta t}.
\end{align}
\\
The Prop.~\ref{prop1bid}.\ref{DiffPbidprop} can be proved using--
\begin{subequations}
    \begin{equation}
\label{eq:DelPL}
\begin{aligned}
       &\nabla_{\bm{P}}L^* = \nabla_{\bm{P}}C_\mathrm{IDSO}(\bm{P^*}) - \bm{C}\bm{\lambda}^{p^{*T}} + 2\bm{D}_r^TC^{-T}[(\overline{\bm{\mu}}^{{v}^*})^T \\&- (\underline{\bm{\mu}}^{{v}^*})^T] +\sum_{e \in \mathcal{E}^P}[\beta_e(\bm{\mu}^{{P}^*}(e))^T] + \sum_{e \in \mathcal{E}^{sub}}[\beta_e\bm{C}_0 (\bm{\mu}^{{sub}^*}(e))^T],\notag
    \end{aligned}
\end{equation}
which can be rearranged as:
\begin{equation}
\begin{aligned}
      \bm{C}\bm{\lambda}^{p^{*T}} &= \nabla_{\bm{P}}C_\mathrm{IDSO}(\bm{P^*}) + 2\bm{D}_r^TC^{-T}[(\overline{\bm{\mu}}^{{v}^*})^T - (\underline{\bm{\mu}}^{{v}^*})^T] \\&+\sum_{e \in \mathcal{E}^P}[\beta_e(\bm{\mu}^{{P}^*}(e))^T] + \sum_{e \in \mathcal{E}^{sub}}[\beta_e\bm{C}_0 (\bm{\mu}^{{sub}^*}(e))^T].\notag
\end{aligned}
\end{equation}
\end{subequations}
The Prop.~\ref{prop1bid}.\ref{DiffQbidprop} can be proved using--
\begin{subequations}
    \begin{equation}
    \label{eq:DelQL}
    \begin{aligned}
        \nabla_{\bm{Q}}L^* = \,&- \bm{C}\bm{\lambda}^{q^{*T}} + 2\bm{D}_x^TC^{-T}[(\overline{\bm{\mu}}^{{v}^*})^T - (\underline{\bm{\mu}}^{{v}^*})^T] \\&+ \sum_{e \in \mathcal{E}^P}[\delta_e(\bm{\mu}^{{P}}(e))^T] + \sum_{e \in \mathcal{E}^{sub}}[\delta_e\bm{C}_0 (\bm{\mu}^{sub}(e))^T],\notag
    \end{aligned}
\end{equation}
which can be rearranged as:
\begin{equation}
\begin{aligned}
     \bm{C}\bm{\lambda}^{q^{*T}} &= 2\bm{D}_x^TC^{-T}[(\overline{\bm{\mu}}^{{v}^*})^T - (\underline{\bm{\mu}}^{{v}^*})^T] \\&+ \sum_{e \in \mathcal{E}^P}[\delta_e(\bm{\mu}^{{P}^*}(e))^T] + \sum_{e \in \mathcal{E}^{sub}}[\delta_e\bm{C}_0 (\bm{\mu}^{{sub}^*}(e))^T].\notag
\end{aligned}
\end{equation}
\end{subequations}
\end{IEEEproof}
\section{Proof of Proposition~\ref{offerprop}}
\label{Append:prop2}
\begin{IEEEproof}
    The proof is similar to the one in Appendix~\ref{Append:Prop1}. Firstly, formulate the Lagrangian as in \eqref{eq:lagrangian} and continue similarly to equation \eqref{eq:gamma_psi}, given as,
\begin{equation}
    \begin{aligned}
        \gamma(\psi) = \frac{\sum_{\phi \in \bm{\Phi}(\psi)}-\lambda_{i,\phi}^p + \eta(\psi) \sum_{\phi \in \bm{\Phi}_\psi} -\lambda_{i,\phi}^q}{N_{\phi_\psi} S_\mathrm{base}\Delta t}.
    \end{aligned}
\end{equation}
As $\psi$ is now an offering DER, $\gamma(\psi) = \pi(\psi) - \frac{M}{p(\psi)}$, the previous equation becomes,
\begin{equation}
        \begin{aligned}
        \pi(\psi) =\frac{M}{p(\psi)} + \frac{\sum_{\phi \in \bm{\Phi}(\psi)}-\lambda_{i,\phi}^p + \eta(\psi) \sum_{\phi \in \bm{\Phi}_\psi} -\lambda_{i,\phi}^q}{N_{\phi_\psi} S_\mathrm{base}\Delta t}; \notag
    \end{aligned}
\end{equation}
An offering DER would produce only when its offer price is less than or equal to the marginal value of procuring power at the connected location.
Therefore, the following equation holds for offers:
\begin{equation}
    \begin{aligned}
        \pi(\psi) \leq \frac{M}{p(\psi)} + \frac{\sum_{\phi \in \bm{\Phi}(\psi)}-\lambda_{i,\phi}^p + \eta(\psi) \sum_{\phi \in \bm{\Phi}_\psi} -\lambda_{i,\phi}^q}{N_{\phi_\psi} S_\mathrm{base}\Delta t}.
    \end{aligned}
\end{equation}
Hence, the qualification price is set as
\begin{align}
    \pi^\mathrm{QP}(\psi) \coloneq \frac{M}{p(\psi)} + \frac{\sum_{\phi \in \bm{\Phi}(\psi)}-\lambda_{i,\phi}^p + \eta(\psi) \sum_{\phi \in \bm{\Phi}_\psi} -\lambda_{i,\phi}^q}{N_{\phi_\psi} S_\mathrm{base}\Delta t}.
\end{align}
\end{IEEEproof}
\section{Proof of Proposition~\ref{eitherprop}}\label{Append:prop3}
\begin{IEEEproof}
    Similar to Appendices~\ref{Append:Prop1} and \ref{Append:prop2}. Since the results are kept in separate bins, Bin A and Bin B. The Qualification price for bids and offers utilises the dual variables in that bin; therefore, for bids in Bin A, the qualification price is set as, 
    \begin{align}
        \pi^\mathrm{QP}(\psi) \coloneq \frac{\sum_{\phi \in \bm{\Phi}(\psi)}-\lambda_{A,i,\phi}^p + \eta(\psi) \sum_{\phi \in \bm{\Phi}_\psi} -\lambda_{A,i,\phi}^q}{N_{\phi_\psi} S_\mathrm{base}\Delta t}.
    \end{align}
    And for offers in Bin B, the qualification price is set as,
    \begin{align}
        \pi^\mathrm{QP}(\psi) \coloneq \frac{M}{p(\psi)} + \frac{\sum_{\phi \in \bm{\Phi}(\psi)}-\lambda_{B,i,\phi}^p + \eta(\psi) \sum_{\phi \in \bm{\Phi}_\psi} -\lambda_{B,i,\phi}^q}{N_{\phi_\psi} S_\mathrm{base}\Delta t}
    \end{align}
\end{IEEEproof}
\bibliographystyle{IEEEtran}
\bibliography{paper}
\end{document}